\documentclass[12pt]{article}
\pdfoutput=1
\usepackage{amsmath,amsfonts,graphicx,color,bbm,tikz}
\usepackage{amssymb}
\usepackage{comment}
\usepackage[nosort]{cite}
\usepackage{subfigure}
\usetikzlibrary{calc,positioning}
\usetikzlibrary{patterns,arrows,decorations.pathreplacing}
\usepackage{caption}
\usepackage{ulem}
\tikzset{>=stealth}

\newcommand{\be}{\begin{equation}}
\newcommand{\ee}{\end{equation}}
\newcommand{\bea}{\begin{eqnarray}}
\newcommand{\eea}{\end{eqnarray}}
\newcommand{\Tr}{{\rm Tr}}

\newcommand{\rF}{{\rm F}}
\newcommand{\rM}{{\rm M}}

\newcommand{\bra}[1]{{\left< {#1} \right|}}
\newcommand{\ket}[1]{{\left| {#1} \right>}}

\def\nn{\nonumber}

\newcommand{\binomi}[2]{\begin{pmatrix} #1 \\ #2 \end{pmatrix}}

\renewcommand{\title}[1]{\vbox{\center\LARGE{#1}}\vspace{3mm}}
\renewcommand{\author}[1]{\vbox{\center#1}\vspace{3mm}}

\newcommand{\email}[1]{\vbox{\center\tt#1}\vspace{3mm}}

\begin{document}

\rightline{\small{\tt }}
\begin{center}

\vskip-1.5cm
{\large {\bf R\'enyi entropy of highly entangled spin chains} }
\vskip 0.75cm

 Fumihiko Sugino$^{*}$ and Vladimir Korepin$^\dagger$

\vskip 0.5cm 
${}^*$~Fields, Gravity \& Strings Group, Center for Theoretical Physics of the Universe, 
Institute for Basic Science (IBS), 55, Expo-ro, Yuseong-gu, Daejeon 34126, Republic of Korea\\
${}^\dagger$~C.N.Yang Institute for Theoretical Physics, Stony Brook University, NY 11794, USA
\email{fusugino@gmail.com, korepin@gmail.com}

\vskip 0.5cm 

\end{center}


\abstract{
\noindent 
Entanglement is one of the most intriguing features of quantum theory and a main resource in quantum information science. 
Ground states of quantum many-body systems with local interactions typically obey an ``area law'' meaning the entanglement entropy proportional to the boundary length. 
It is exceptional when the system is gapless, and the area law had been believed to be violated by at most a logarithm for over two decades. 
Recent discovery of Motzkin and Fredkin spin chain models is striking, since these models provide significant violation of the entanglement beyond the belief, growing as a square root of the volume in spite of local interactions. 
Although importance of intensive study of the models is undoubted to reveal novel features of quantum entanglement, it is still far from their complete understanding.  
In this article, we first analytically compute the R\'enyi entropy of the Motzkin and Fredkin models by careful treatment of asymptotic analysis. 
The R\'enyi entropy is an important quantity, since the whole spectrum of an entangled subsystem is reconstructed once the R\'enyi entropy is known as a function of its parameter.      
We find non-analytic behavior of the R\'enyi entropy with respect to the parameter, which is a novel phase transition never seen in any other spin chain studied so far. 
Interestingly, similar behavior is seen in the R\'enyi entropy of Rokhsar-Kivelson states in two-dimensions.
}




\newpage

\section{Introduction}
Entanglement is the most distinguished feature of quantum systems, and a core of quantum computation and quantum information theory~\cite{bennett_etal,nielsen-chuang}. 
Recently, further attention has been payed to the entanglement in quantum field theory (for example see Ref.~\cite{witten}). 
When a subsystem $A$ is picked from a given full system $S$, entanglement between $A$ and its complement $B=S-A$ is extracted from the reduced density matrix of $A$: $\rho_A$, 
which is obtained by tracing out the density matrix of the full system $\rho$ 
with respect to the Hilbert space belonging to $B$: 
$
\rho_A={\rm Tr}_B\,\rho
$. 
When $\rho$ is a pure state, the amount of the entanglement is normally measured by the von Neumann entanglement entropy 
$
S_A=-{\rm Tr} \left(\rho_A\ln \rho_A\right)
$. 
However, the R\'enyi entropy \cite{renyi_book} defined by 
\begin{equation}
S_{A,\,\alpha}=\frac{1}{1-\alpha}\,\ln {\rm Tr}\, \rho_A^\alpha
\label{renyiA}
\end{equation}
with $\alpha$ real positive not equal to 1 has further importance, 
because we can reconstruct the whole spectrum (entanglement spectrum) of $\rho_A$ or equivalently of the entanglement Hamiltonian 
$
H_{{\rm ent},\,A}\equiv -\ln \rho_A
$ 
once $S_{A,\,\alpha}$ is known as a function of $\alpha$. By taking the limit $\alpha\to 1$, the R\'enyi entropy reduces to the von Neumann entanglement entropy.  
We can represent the reduced density matrix in terms of the entanglement Hamiltonian as 
$\rho_A= \exp(-H_{{\rm ent},\,A})$. 
Then $\Tr_A\,\rho_A^\alpha$ is a partition function of this Hamiltonian and the R\'enyi entropy (\ref{renyiA}) is proportional to free energy of the entanglement Hamiltonian. 
The $\alpha$ plays the role of inverse temperature. In this paper, we discover a phase transition with respect to this temperature.  

Ground states of quantum many-body systems typically show an ``area law'' that means the amount of entanglement between the subsystems $A$ and $B$ with $\rho$ being the density matrix of the ground states 
is proportional to the area of the boundaries of $A$ and $B$. 
More precisely, for a quantum many-body system with local interactions on $D$-dimensional spatial lattice, its von Neumann entanglement entropy for the ground state is believed to behave as 
$
S_A=O\left(L^{D-1}\right)
$
when the system has a gap. 
$L$ is a typical length scale of the subsystem $A$, and the behavior is estimated for $L$ large.  
On the other hand, when the system is gapless, it is expected to obey similar behavior but with a possible logarithmic correction as 
$
S_A=O\left(L^{D-1}\ln L\right)
$
~\cite{eisert_Rev}. 
For example, an exactly solvable spin chain introduced by Affleck, Kennedy, Lieb and Tasaki (the AKLT model)~\cite{AKLT1,AKLT2} has a gapped spectrum 
and $S_A$ yields a constant value independent of $L$, which coincides with the area law for $D=1$~\cite{fan}. 
Other spin chains (XXZ and XY models) which contain critical regime realizing gapless spectrum are investigated to find consistency with the above behavior~\cite{vidal,jin}. 
Results of $(1+1)$D conformal field theories (CFTs)~\cite{wilczek,korepin,calabrese} and of the Fermi liquid theory~\cite{wolf} also yield the logarithmic correction with $D=1$.  
The rigorous proof of the belief and expectation has not been provided yet, except for the area law for gapped 1D systems: $S_A=O(L^0)$~\cite{hastings}~\footnote{
A proof of the area law of the R\'enyi entropy for gapped 1D systems is given in \cite{huang}.
}.

\section{Motzkin and Fredkin Spin Chains}
\begin{figure}[h!]
\captionsetup{width=0.8\textwidth}
\centering
\includegraphics[width=0.9\textwidth]{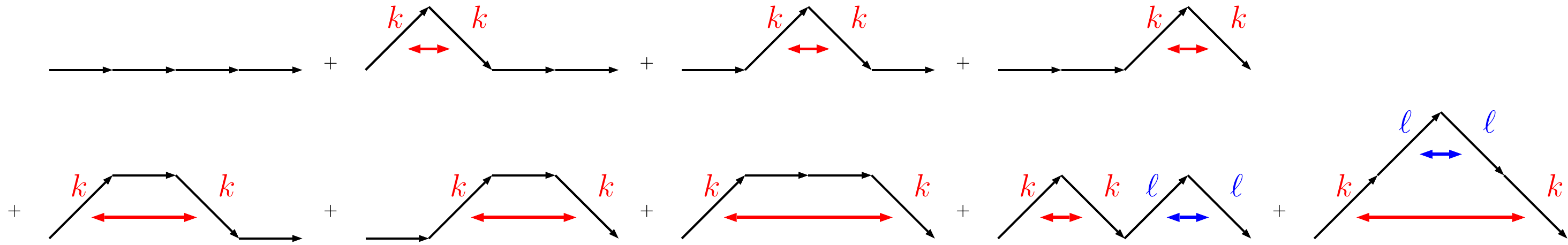}
\caption{\small Colored Motzkin walks of length $2n=4$. Up- and down-steps with the same color are matched. Colored Dyck walks consist of the last two terms.}
\label{fig:CMW4}
\end{figure}

This situation has been drastically changing, since a solvable spin chain model was introduced by Movassagh and Shor~\cite{motzkin} (this model is called as Motzkin spin chain). 
Spin degrees of freedom at each lattice site are $s$ kinds of up- and down-spins ($|u^k\rangle$ and $|d^k\rangle$ with $k\in\{1,2,\cdots, s\}$), 
and a single zero spin ($|0\rangle$), which compose spin-$s$ degrees of freedom ($2s+1$) with $s=1,2,\cdots$.  
The Hamiltonian of the model consists of nearest neighbor local interactions of the spins, and has a translational invariant bulk part. 
The system is frustration-free and has a unique ground state, whose von Neumann entanglement entropy scales as $\sqrt{L}$ exhibiting extraordinary violation of the area law for $s\geq 2$. 
The ground state is expressed as the equal-weight superposition of an exponentially large number of states with respect to the length of the chain $2n$. 
For $s=1$, each of the states corresponds to a path of random walks in $(x,y)$ plane called as Motzkin walks, 
which consists of up-, down- and flat-steps pointing $(1,1)$, $(1,-1)$ and $(1,0)$ respectively, starts at the origin, ends at $(2n, 0)$, 
and paths are not allowed to enter $y<0$ region. 
In this case, the von Neumann entanglement entropy for the subsystems $A$ and $B$ divided at the middle of the chain behaves as 
$
S_A=\frac12\ln n+ \mbox{(constant)}
$ 
up to terms vanishing as $n\to \infty$~\cite{bravyi_etal}. 
For $s>1$, $s$-color degrees of freedom are assigned by putting a color index $k \in\{1,2,\cdots, s\}$ to up- and down-steps, 
and the color of each up-step should be matched with that of subsequent down-step at the same height in each path. 
Motzkin walks implementing these natures are referred as colored Motzkin walks. As an example, colored Motzkin walks of length $2n=4$ are depicted in Fig.~\ref{fig:CMW4}. 
The matching of colors leads to strong correlation between distant spins achieving the large entanglement of $\sqrt{n}$.   

Salberger and one of the authors (V.K.) constructed an analogous model for half-integer spins~\cite{fredkin} (see also \cite{dellanna_etal}), which is called as Fredkin spin chain after a relation to Fredkin (Controlled -Swap) gates.  
This model has $s$ kinds of up- and down-spins ($|u^k\rangle$ and $|d^k\rangle$ with $k\in\{1,2,\cdots, s\}$) at each site, composing spin-$(s-1/2)$ degrees of freedom. 
Interactions of the Hamiltonian are local ranging up to next-to-nearest neighbors. The ground state is unique, corresponding to random walks called as Dyck walks, in which paths containing a flat step are removed from 
the Motzkin walks. 
For example, colored Dyck walks of length $2n=4$ are the last two terms in Fig.~\ref{fig:CMW4}. 
For some applications in mathematical physics for Dyck paths, see \cite{alcarez,chico}.     
Asymptotic behavior of the von Neumann entanglement entropy is similar to the Motzkin spin chain for colored ($s>1$) as well as uncolored ($s=1$) cases. 
We present details of these two models in Appendices~\ref{sec:GS_Fredkin} and \ref{sec:GS_Motzkin}. 

Deformation of the Motzkin and Fredkin spin chains are discussed to realize further extensive entanglement proportional to the volume in~\cite{zhang_k,salberger_etal,zhang_ak,udagawa_k,barbiero_etal}. 
A parameter $t$ is introduced in the deformation, and ground states are expressed by a superposition of the same states as the original but now with the weight factor $t^{\cal A}$ to each state, 
where ${\cal A}$ denotes the area bounded by the corresponding path and the $x$-axis. For colored case ($s>1$) with $t>1$, paths reaching height of $O(n)$ dominantly contribute to the von Neumann entanglement entropy, 
leading to linear scaling in $n$. 
From a different point of view, an extension of the models by symmetric inverse semigroups is discussed in~\cite{SISmotzkin,SISfredkin}. 
There are excited states corresponding to disconnected paths, for which localization phenomena occur. 
In addition, a new spin chain called the pair-flip model is introduced in~\cite{caha_n}, where the Hamiltonian is local, translational invariant and does not have boundary terms. 
Since the Hamiltonian is not frustration-free, fully analytic treatment is hard, but it is conjectured from partial analytic derivation and numerical results that the ground state is unique and its entanglement entropy 
obeys the same scaling of a square root of the volume as the Motzkin and Fredkin models.  

The Motzkin and Fredkin spin chains are critical but cannot be described by relativistic CFTs due to the gap scaling as $n^{-z}$ with $z\geq 2$. 
Although some two-point correlation functions have been computed in colorless cases~\cite{dellanna_etal,movassagh}, further investigation needs to reveal dynamical properties of the models, in particular for colored cases.      
 
\begin{figure}[h!]
\captionsetup{width=0.8\textwidth}
\centering
\includegraphics[width=.9\linewidth]{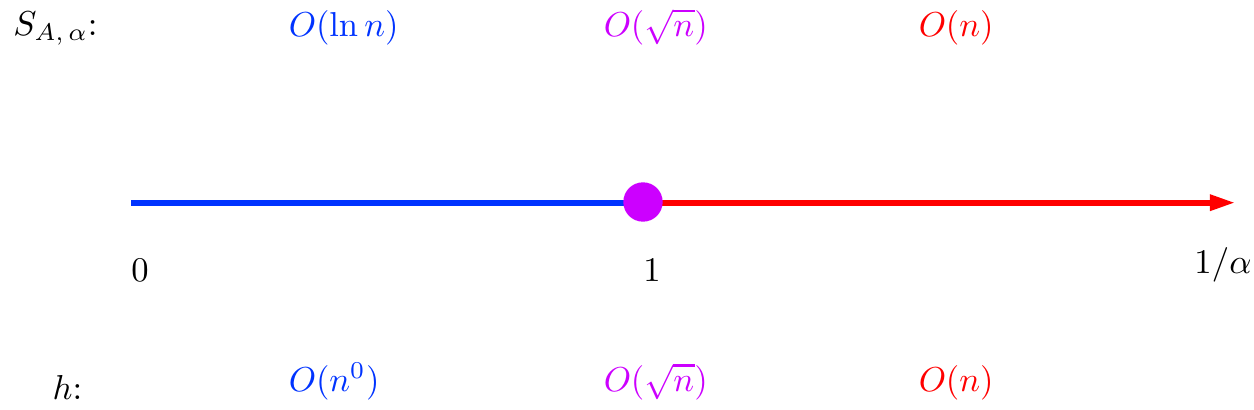}
\caption{\small Phase diagram for the R\'enyi entropy in Motzkin and Fredkin spin chains with respect to $\alpha$. In writing the reduced density matrix in terms of the entanglement Hamiltonian as $\rho_A=e^{-H_{{\rm ent},\,A}}$, 
$1/\alpha$ can be interpreted as temperature. $S_{A,\,\alpha}=O(n)$ for \textcolor{red}{$0<\alpha<1$ (``high temperature'')}, whereas $S_{A,\,\alpha}=O(\ln n)$ for \textcolor{blue}{$\alpha>1$ (``low temperature'')}. 
At the transition point, the von Neumann entanglement entropy behaves as $O(\sqrt{n})$. At the bottom, height of paths dominantly contributing to the R\'enyi entropy is shown.}
\label{fig:phase}
\end{figure}

In this article, we analytically compute the R\'enyi entropy in the Motzkin and Fredkin models in colored as well as colorless cases by careful treatment of asymptotic analysis. 
Because the R\'enyi entropy has sufficient information to reconstruct the whole entanglement spectrum, the result is crucial to understand the full of entanglement property of the models. 
Actually, we find a phase transition at $\alpha=1$ in the colored cases. Namely, (\ref{renyiA}) has different asymptotic behavior for $0<\alpha<1$ and $\alpha>1$ as depicted in Fig.~\ref{fig:phase}. 
The former scales as a linear of $n$, whereas the latter as $\ln n$. The phase transition point itself forms a phase, in which the von Neumann entanglement entropy scales as $\sqrt{n}$.           
Remarkably, this kind of phase transition has never been found in any other spin chain computed so far. 
It is interesting to note that similar behavior has been discovered in the R\'enyi entropy of Rokhsar-Kivelson states in two dimensions~\cite{stephan}.  
In particular, the transition occurs in quantum dimer state on the square lattice at the same value of the R\'enyi parameter $\alpha=1$. 
Also, similar phase transition in the holographic context has been reported in Ref.~\cite{belin}, where the transition point is strictly larger than 1. 

\section{Mini Review of R\'enyi Entropy in Quantum Spin Chains}
\setcounter{equation}{0}
In this section, we summarize results of the R\'enyi entropy in quantum spin chains obtained so far. 
\subsection*{AKLT model}
The AKLT model consists of a linear chain of $N$ spin-1's in the bulk ($\vec{S}_j$ at sites $j=1,2,\cdots, N$) and two spin-$1/2$'s on the boundary ($\vec{s}_b$ at $b=0,N+1$). 
The Hamiltonian is 
\begin{equation}
H_{{\rm AKLT}}= \sum_{j=1}^{N-1}\left[\vec{S}_j\cdot\vec{S}_{j+1} + \frac13\left(\vec{S}_j\cdot\vec{S}_{j+1}\right)^2 \right]+\pi_{0,1}+\pi_{N,N+1},
\label{H_AKLT}
\end{equation}
where the boundary terms $\pi_{0,1}=\frac23\left(1+\vec{s}_0\cdot\vec{S}_1\right)$ and $\pi_{N,N+1}=\frac23\left(1+\vec{s}_N\cdot\vec{S}_{N+1}\right)$ are projectors onto spin-$3/2$ states. 
The ground state $|{\rm GS}\rangle$ is unique, and the spectrum is gapped~\cite{AKLT1,AKLT2}. 
Let us start with the full density matrix of the ground state $\rho=|{\rm GS}\rangle\langle {\rm GS}|$. 
We pick $L$ contiguous spin-1's at sites $k, k+1, \cdots, k+L-1$ as a subsystem $A$ and trace out all the other spins including the boundary spins to obtain the reduced density matrix of $A$: 
\begin{equation}
\rho_A={\rm Tr}_{0,1,\cdots,k-1,k+L, K+L+1,\cdots,N,N+1}\,\rho. 
\label{rhoA_AKLT}
\end{equation}
The subscripts of the symbol ``Tr'' stand for sites of spins to be traced-out. 
It turns out that (\ref{rhoA_AKLT}) does not depend on either $k$ or $N$~\cite{fan}. 
The von Neumann entanglement entropy and the R\'enyi entropy take the same value $2\ln2$ as $L\to \infty$,  
saturating an $L$-independent value. 
Note also that the R\'enyi entropy is independent of the parameter $\alpha$.  
For a general spin $S(=1,2,\cdots)$ version of the AKLT model, the result is similar with the value $2\ln (S+1)$~\cite{ying_etal}.    

\subsection*{XY model}
The XY model of an infinite chain in a transverse magnetic field is given by the Hamiltonian:
\begin{equation}
H_{{\rm XY}}= -\sum_{j=-\infty}^\infty \left[(1+\gamma) \sigma^x_j\sigma^x_{j+1} + (1-\gamma) \sigma^y_j\sigma^y_{j+1} + 2h\sigma^z_j\right],
\label{H_XY}
\end{equation} 
where $\sigma^x_j, \sigma^y_j, \sigma^z_j$ are Pauli matrices describing spin operators at the site $j$. $\gamma\geq 0$ is the anisotropy parameter, and $2h\geq 0$ is the magnetic field. 
This model with zero and nonzero $h$ was solved in~\cite{lieb_sm,barouch_md,barouch_m}. 
It can be mapped to a system of free fermions with a spectrum 
\begin{equation}
\epsilon_k=4\sqrt{(\cos k-h)^2+\gamma^2\sin^2k}, 
\label{e_XY}
\end{equation}
which becomes gapless for $(h,\gamma)=(h,0)$ with $0\leq h\leq 1$ or $(h,\gamma)=(1, \gamma)$. 
The former case reduces to isotropic XY model or XX model in a magnetic field, and the latter case means that the critical magnetic field is $2h_c=2$. 
The system is off critical in any other region of the $(h,\gamma)$-plane. 
The line of $\gamma=1$ also corresponds to the quantum Ising chain in a magnetic field. 

We pick a block of $L$ neighboring spins as a subsystem $A$, and compute the entanglement of $A$ from the density matrix of the ground state.  
By introducing elliptic parameters:
\begin{equation}
k=\begin{cases} \frac{1}{\gamma}\sqrt{h^2+\gamma^2-1}, & \mbox{for } 1-\gamma^2<h^2<1 \\
                          \sqrt{\frac{1-h^2-\gamma^2}{1-h^2}}, & \mbox{for } h^2< 1-\gamma^2 \\
                          \frac{\gamma}{\sqrt{h^2+\gamma^2-1}}, & \mbox{for } h>1,
\end{cases}
\end{equation} 
$k'=\sqrt{1-k^2}$ and $q=\exp\left(-\pi\frac{I(k')}{I(k)}\right)$ with the complete elliptic integral of the first kind 
\begin{equation}
I(k)=\int_0^1\frac{dx}{\sqrt{(1-x^2)(1-k^2x^2)}},
\end{equation} 
the R\'enyi entropy in $L\to \infty$ limit is expressed as 
\begin{align}
S_{A,\,\alpha}= & \, \frac16\frac{\alpha}{1-\alpha}\ln(kk')-\frac13\frac{1}{1-\alpha}\ln\left(\frac{\theta_2(0,q^\alpha)\,\theta_4(0,q^\alpha)}{\theta_3^2(0,q^\alpha)}\right) \nonumber \\
& +\frac13\ln 2
\label{XY_renyi1}
\end{align}
for $h>1$ and 
\begin{align}
S_{A,\,\alpha}= & \, \frac16\frac{\alpha}{1-\alpha}\ln(\frac{k'}{k^2})-\frac13\frac{1}{1-\alpha}\ln\left(\frac{\theta^2_2(0,q^\alpha)}{\theta_3(0,q^\alpha)\,\theta_4(0,q^\alpha)}\right) \nonumber \\
& +\frac13\ln 2
\label{XY_renyi2}
\end{align}
for $h<1$~\cite{franchini_ik}. $\theta_j(z,q)$ ($j=1,2,3,4$) are the Jacobi theta functions. 
(\ref{XY_renyi1}) and (\ref{XY_renyi2}) are automorphic functions of $\alpha$, and can take any positive value. 
In $\alpha\to 1$ limit, these expressions reduce to the von Neumann entanglement entropy obtained in~\cite{jin,its_jk,peschel} :
\begin{align}
S_A=& \, \frac16\left[\ln \frac{4}{kk'} + (k^2-k'^2)\frac{2I(k)I(k')}{\pi}\right], \qquad (h>1); 
\label{XY_EE1}
\\
 S_A=& \, \frac16\left[\ln \frac{4k^2}{k'} + (2-k^2)\frac{2I(k)I(k')}{\pi}\right], \qquad (h<1).
 \label{XY_EE2}
\end{align}
For generic point in $(h,\gamma)$ plane, the expressions (\ref{XY_renyi1})--(\ref{XY_EE2}) are finite constants. 

In the critical case of the XX model $\gamma\to 0$ with $h<1$, (\ref{XY_renyi2}) and (\ref{XY_EE2}) diverge as  
\begin{align}
S_{A,\,\alpha} = & \, \frac{1+\alpha}{\alpha}\left(-\frac16\ln\gamma+\frac{1}{12}\ln(1-h^2)+\frac13\ln 2\right), 
\label{XY_renyi3}
\\
S_A=& \, -\frac13\ln\gamma+\frac16\ln(1-h^2)+\frac23\ln 2
\label{XY_EE3}
\end{align}
up to terms of $O(\gamma^2\ln\gamma)$. 
These can be compared with results obtained in keeping the block size $L$~\cite{jin}:
\begin{align}
S_{A,\,\alpha}\approx & \begin{cases} \frac{1}{1-\alpha}\ln\left[\left(\frac{{\cal L}}{2\pi}\right)^\alpha+\left(1-\frac{{\cal L}}{2\pi}\right)^\alpha\right], & (0<{\cal L} \ll 1); \\
\frac{1+\alpha}{6\alpha}\ln{\cal L} +\Upsilon^{\{\alpha\}}_1, & ({\cal L}\gg 1),
\end{cases}
\label{XX_renyi}
\\
S_A\approx & \begin{cases} \frac{{\cal L}}{2\pi}\ln \frac{2\pi}{{\cal L}} +O({\cal L}), & (0<{\cal L}\ll 1);\\
\frac13\ln{\cal L}+O(1), & ({\cal L}\gg 1)
\end{cases}
\label{XX_EE}
\end{align}
with a universal scaling variable ${\cal L}=2L\sqrt{1-h^2}$ and an $\alpha$-dependent constant $\Upsilon^{\{\alpha\}}_1$.

The other critical case $h\to 1$ with $\gamma\neq 0$ reduces (\ref{XY_renyi1})--(\ref{XY_EE2}) to 
\begin{align}
S_{A,\,\alpha} =& \, \frac{1+\alpha}{\alpha}\left(-\frac{1}{12}\ln |h-1| +\frac16\ln\gamma+\frac14\ln 2\right), 
\label{XY_renyi4}
\\
S_A=& \, -\frac16\ln |h-1|+\frac13\ln \gamma +\frac12\ln 2
\label{XY_EE4}
\end{align}
up to $O(|h-1|\ln |h-1|)$ terms. The Ising model belongs to this case with $\gamma=1$. 
Note that $\ln |h-1|$ in (\ref{XY_renyi4}) and (\ref{XY_EE4}) can be written as $-\ln\xi$ in terms of the correlation length $\xi$ from~\cite{barouch_m}. 

We see the pole at $\alpha=0$ in the above R\'enyi entropies (\ref{XY_renyi3}), (\ref{XX_renyi}) and (\ref{XY_renyi4}). 
This reflects the fact that the dimension of the Hilbert space of $A$ becomes infinite in the $L\to\infty$ limit. 
These are comparable to the R\'enyi entropy of $(1+1)$D CFT~\cite{wilczek,perlmutter}:
\begin{equation}
S_{A,\,\alpha}=\frac{c_L+c_R}{12}\left(1+\frac{1}{\alpha}\right)\ln\frac{L}{\epsilon},
\label{CFT_renyi}
\end{equation}
where $c_L$ and $c_R$ are left and right central charges, and $\epsilon$ is an ultraviolet cutoff. 
(\ref{CFT_renyi}) is consistent with the critical regime of the XX model described by a free boson CFT of $c_L=c_R=1$ 
and that of the Ising model by a free fermion CFT of $c_L=c_R=1/2$.    

\subsection*{Colorless Motzkin model}
The R\'enyi entropy in the colorless ($s=1$) Motzkin spin chain~\cite{bravyi_etal} is computed in~\cite{movassagh} for two choices of a subsystem $A$. 
One of the choices is the first contiguous $n_1$ spins of the length-$2n$ chain. The result for $n_1=O(n)\to \infty$ reads 
\begin{align}
S_{A,\,\alpha}= &\, \frac12 \ln \frac{n_1(2n-n_1)}{n}+\frac{1}{1-\alpha}\ln\Gamma\left(\alpha+\frac12\right) \nonumber \\
& -\frac{1}{2(1-\alpha)} \left\{(1+2\alpha)\ln \alpha + \alpha\ln \frac{\pi}{24} +\ln 6\right\}.
\label{s1motzkin_renyi1}
\end{align}
The other choice is the $L$-consecutive spins 
centered in the middle of length-$2n$ chain. 
For $1\ll L\ll n$, the R\'enyi entropy behaves as 
\begin{equation}
S_{A,\,\alpha} = \frac12\ln L + \ln \left(2\sqrt{\frac{\pi}{3}}\right) -\frac{\ln \alpha}{2(1-\alpha)}.
\label{s1motzkin_renyi2}
\end{equation}
The leading term has also been obtained from a continuum version of the ground-state wavefunction\cite{chen} that also gives the same leading behavior 
in the colorless Fredkin spin chain.  
These two expressions share the logarithmic growth with the size of the subsystem $A$ and a logarithmic singularity at $\alpha=0$ in the 
correction terms.    

\subsection*{Deformed Fredkin model}
For the deformed Fredkin spin chain by a parameter $t>1$~\cite{salberger_etal}, the R\'enyi entropy is obtained in~\cite{udagawa_k}, where a subsystem $A$ consists of the first block of $n_1$ spins of the length-$2n$ chain. 
For both cases of $0<\alpha<1$ and $\alpha>1$, the R\'enyi entropy behaves as
\begin{equation}
S_{A,\,\alpha}=n_1\ln s + O(1) .
\label{dfredkin_renyi}
\end{equation}  
It is proportional to the volume of the subsystem $A$ for the colored case $s>1$. 
(\ref{dfredkin_renyi}) coincides with the expression of the von Neumann entanglement entropy.

\section{R\'enyi Entropy of Fredkin Spin Chain}
\setcounter{equation}{0}
The ground state of the Fredkin spin chain of length $2n$ is expressed as the uniform superposition of states, each of which 
corresponds to a Dyck path from $(0,0)$ to $(2n,0)$. The number of all possible Dyck paths is given by 
\begin{equation}
N_{{\rm F},\,2n,\,s}=s^nN_{{\rm F},\,2n}=\frac{s^n}{n+1} \begin{pmatrix}2n \\ n \end{pmatrix},
\label{NF}
\end{equation} 
which grows exponentially with respect to $n$. This appears as a normalization factor of the ground state:
\begin{equation}
|{\rm GS}\rangle=\frac{1}{\sqrt{N_{{\rm F},\,2n,\,s}}} \sum_{w\in \{\textrm{length-$2n$ Dyck paths}\}} |w\rangle.
\end{equation}
We divide the full system into left and right half chains $A$ and $B$, and compute the R\'enyi entropy of $A$. 
Let us consider a part of Dyck paths belonging to $A$, which consists of paths from $(0,0)$ to $(n,h)$. 
Height $h$ takes nonnegative integers. Each of them contains $h$ unmatched up-steps. 
The number of such paths is 
\begin{equation}
N^{(0\to h)}_{{\rm F},\,n,\,s}=s^{\frac{n+h}{2}}N^{(h)}_{{\rm F},\,n} 
=s^{\frac{n+h}{2}}\frac{1+(-1)^{n+h}}{2}\frac{h+1}{\frac{n+h}{2}+1} \begin{pmatrix} n \\ \frac{n+h}{2} \end{pmatrix}. 
\label{NFh}
\end{equation}
As a result of the Schmidt decomposition, the R\'enyi entropy takes the form (see Appendix~\ref{sec:GS_Fredkin} for the derivation):
\begin{equation}
S_{{\rm F},\,A,\alpha}=\frac{1}{1-\alpha}\ln\sum_{h=0}^ns^h\left(p^{(h)}_{{\rm F},\,n,n,\,s}\right)^\alpha,
\label{renyiF}
\end{equation}
where the factor $s^h$ comes from matching of colors across the boundary of $A$ and $B$, 
and the probability factor is given by (\ref{NF}) and (\ref{NFh}) as 
\begin{equation}
p^{(h)}_{{\rm F},\,n,n,\,s}\equiv s^{-h}\frac{\left(N^{(h)}_{{\rm F},\,n}\right)^2}{N_{{\rm F},\,2n}}.
\label{pnnF}
\end{equation}
It is straightforward to see the von Neumann entanglement entropy is written as 
\begin{equation}
S_{{\rm F},\,A}=- \sum_{h=0}^n s^h p^{(h)}_{{\rm F},\,n,n,\,s} \ln p^{(h)}_{{\rm F},\,n,n,\,s}.
\label{vNEEF}
\end{equation}
For $n, n\pm h\gg 1$, (\ref{pnnF}) behaves as 
\begin{align}
p^{(h)}_{{\rm F},\,n,n,\,s}\simeq  & \,\frac{1+(-1)^{n+h}}{2}\frac{8}{\sqrt{\pi}}\,s^{-h}\frac{(h+1)^2}{n^{3/2}} \nonumber \\
&\,\times \left(\frac{n}{n+h}\right)^{n+h+3} \left(\frac{n}{n-h}\right)^{n-h+1}\times \left[1+O(n^{-1})\right]. 
\label{pnnFasym}
\end{align}
In case of $h\leq O(n^{1/2})$, this is further simplified to yield 
\begin{align}
p^{(h)}_{{\rm F},\,n,n,\,s} \simeq & \, 
\frac{1+(-1)^{n+h}}{2}\,\frac{8}{\sqrt{\pi}}\,s^{-h}\frac{(h+1)^2}{n^{3/2}}\,e^{-\frac{(h+1)^2}{n}} \nonumber \\
& \times \left[1+O(n^{-1})\right].
\label{pnnFasym2}
\end{align}
Notice that we can use (\ref{pnnFasym2}) only if height $h\leq O(n^{1/2})$ dominantly contributes to the sum (\ref{renyiF}). 
Otherwise, we should compute (\ref{renyiF}) with use of (\ref{pnnFasym}). 
This systematic analysis of $p^{(h)}_{{\rm F},\,n,n,\,s}$ is crucial to obtain the correct asymptotic form of the R\'enyi entropy. 

\subsection*{Colorless case ($s=1$)}
In the colorless case, we can evaluate the sum by using (\ref{pnnFasym2}), and finally obtain 
\begin{align}
S_{{\rm F},\,A,\alpha}=& \, \frac12 \ln n+\frac{1}{1-\alpha}\ln\Gamma\left(\alpha+\frac12\right) \nonumber \\
&-\frac{1}{2(1-\alpha)} \left\{(1+2\alpha)\ln \alpha + \alpha\ln \frac{\pi}{64} +\ln 16\right\}
\label{renyiFcolorless}
\end{align}
up to terms vanishing as $n\to \infty$ (see Appendix~\ref{sec:renyi_F} for the derivation). 
This expression gives the von Neumann entanglement entropy in the limit $\alpha\to 1$: 
\begin{equation}
S_{{\rm F},\,A}=\frac12 \ln n+\frac12\ln \frac{\pi}{4}+\gamma-\frac12.
\label{vNEEFcolorless}
\end{equation}

\subsection*{Colored case ($s>1$)}
In the colored case, we can see that the order of $h$ dominantly contributing to (\ref{renyiF}) changes depending on the value of $\alpha$. 
Note that the summand of (\ref{renyiF}) contains the factor $s^{(1-\alpha)h}$. 
For $0<\alpha<1$, this exponentially grows with $h$, which leads to the saddle point $h_*=O(n)$ in the sum. 
On the other hand, for $\alpha>1$, the factor exponentially decays and height $h\leq O(n^0)$ is dominant in the sum. 
We present detailed analysis and computation in Appendix~\ref{sec:renyi_Fs}. 

The result for \textcolor{red}{$0<\alpha<1$} reads 
\begin{align}
S_{{\rm F},\,A, \alpha} = & \, n\frac{2\alpha}{1-\alpha}\,\ln\cosh\frac{\theta}{2} +\frac{1+\alpha}{2(1-\alpha)}\,\ln n \nonumber \\
& -\ln s +\frac12\ln\frac{\pi}{4} -\frac{1}{2(1-\alpha)}\,\ln \alpha \nonumber \\
&-\frac{1}{1-\alpha}\,\ln\cosh\frac{\theta}{2}
+\frac{2\alpha}{1-\alpha}\,\ln \sinh\theta 
\label{renyiFcolor1}
\end{align}
with $\theta\equiv \frac{1-\alpha}{\alpha}\ln s$.  
The R\'enyi entropy grows proportionally to the volume as $n\to \infty$. 
In contrast to the colorless case and the results of other spin chains, we cannot take $\alpha\to 1$ or $s\to 1$ limit in this expression. 
The limit $\alpha\to 1$ or $s\to 1$ does not commute with $n\to \infty$ limit. 

For \textcolor{blue}{$\alpha>1$}, we obtain 
\begin{align}
S_{{\rm F},\,A, \,\alpha} =& \, \frac{3\alpha}{2(\alpha-1)}\,\ln n +\frac{\alpha}{2(\alpha-1)}\,\ln \frac{\pi}{32^2} \nonumber \\
& -\frac{1}{\alpha-1}\ln \Phi\left(s^{-2(\alpha-1)}, -2\alpha, \frac12\right) 
\label{renyiFcolor2}
\end{align}
for $n$ even, and 
\begin{align}
S_{{\rm F},\,A, \,\alpha} =& \, \frac{3\alpha}{2(\alpha-1)}\,\ln n -\ln s+\frac{\alpha}{2(\alpha-1)}\,\ln \frac{\pi}{32^2} \nonumber \\
& -\frac{1}{\alpha-1}\ln \Phi\left(s^{-2(\alpha-1)}, -2\alpha, 0\right)
\label{renyiFcolor3}
\end{align}
for $n$ odd. 
Here, $\Phi(z,g,a)$ is the Lerch transcendent :
\begin{equation}
\Phi(z,g,a)=\sum_{k=0}^\infty \frac{z^k}{(k+a)^g}.
\label{Lerchtr}
\end{equation}
The R\'enyi entropy has asymptotic behavior of logarithmic growth with $n$. 
The correction terms are different depending on $n$ being even or odd, although the leading logarithmic term is common. 
Again, we cannot take the limit $\alpha\to 1$ or $s\to 1$ in (\ref{renyiFcolor2}) and (\ref{renyiFcolor3}).

\section{R\'enyi Entropy of Motzkin Spin Chain}
\setcounter{equation}{0}
For the Motzkin spin chain of length $2n$, we can also evaluate asymptotic behavior of the R\'enyi entropy in a similar way to the Fredkin model, 
although it is more technically intricate. 
The ground state is the equal-weight superposition of states, which correspond to Motzkin paths from $(0,0)$ to $(2n,0)$. 
The number of paths is given by
\begin{equation}
N_{{\rm M},\,2n,\,s}=\sum_{\rho=0}^n\begin{pmatrix}2n \\ 2\rho \end{pmatrix} s^{n-\rho}N_{{\rm F},\,2n-2\rho}. 
\label{NM}
\end{equation}
A part of the paths belonging to $A$ are paths from $(0,0)$ to $(n,h)$, whose number is 
\begin{equation}
N^{(0\to h)}_{{\rm M},\,n,\,s}=s^h\tilde{N}^{(0\to h)}_{{\rm M},\,n,\,s}=\sum_{\rho=0}^{n-h}\begin{pmatrix} n \\ \rho \end{pmatrix} s^{\frac{n-\rho+h}{2}}N^{(h)}_{{\rm F},\,n-\rho}.
\label{NhM}
\end{equation}
The expression of the R\'enyi entropy of the half chain $A$ is similar to the Fredkin case:
\begin{equation}
S_{{\rm M},\,A,\alpha}=\frac{1}{1-\alpha}\ln\sum_{h=0}^ns^h\left(p^{(h)}_{{\rm M},\,n,n,\,s}\right)^\alpha,
\label{renyiM}
\end{equation}
but the probability $p^{(h)}_{{\rm M},\,n,n,\,s}$ is given by
\begin{equation}
p^{(h)}_{{\rm M},\,n,n,\,s}=\frac{\left(\tilde{N}^{(0\to h)}_{{\rm M},\,n,\,s}\right)^2}{N_{{\rm M},\,2n,\,s}}. 
\label{pnnM}
\end{equation}
We evaluate the sums of (\ref{NM}) and (\ref{NhM}) by the saddle point method for $n, \,\rho, \,n-\rho\pm h\gg 1$, and obtain 
(see Appendix~\ref{sec:GS_Motzkin} for the derivation) 
\begin{align}
 p^{(h)}_{{\rm M}, \, n,n,\,s} \simeq & \, 
\frac{s^{-h}}{\sqrt{\pi}\,s^{1/4}}\,\frac{(2n)^{3/2}}{\left(2\sqrt{s}+1\right)^{2n+\frac32}}\,\frac{n^{2n+1}}{\rho_0^{2n+3}} \nonumber \\
& \times \frac{(h+1)^2}{\left[4sn^2-(4s-1)h^2\right]^{1/2}}\,\left(\frac{n-\rho_0-h}{n-\rho_0+h}\right)^{h+1} \nonumber \\
& \times \left[1+ O(n^{-1})\right],
\label{pnnMasym}
\end{align}
where the saddle point value of $\rho$ is $\rho_0+O(n^0)$ with
\begin{equation}
\rho_0\equiv \frac{n}{4s-1}\left[-1+\sqrt{4s-(4s-1)\frac{h^2}{n^2}}\right].
\end{equation}
When $h\leq O(n^{1/2})$, (\ref{pnnMasym}) reduces to 
\begin{equation}
p^{(h)}_{{\rm M}, \, n,n,\,s} \simeq s^{-h}\sqrt{\frac{2}{\pi\sigma^3}}\,\frac{(h+1)^2}{n^{3/2}}\,e^{-\frac{1}{2\sigma } \frac{(h+1)^2}{n}}
\times \left[1+O(n^{-1})\right]
\label{pnnMasym2}
\end{equation}
with $\sigma\equiv \frac{\sqrt{s}}{2\sqrt{s}+1}$.

\subsection*{Colorless case ($s=1$)}
We can compute (\ref{renyiM}) with (\ref{pnnMasym2}) for the colorless case (see also Appendix~\ref{sec:renyi_M}). The result is 
\begin{align}
S_{{\rm M},\,A,\alpha} = & \, \frac12 \ln n+\frac{1}{1-\alpha}\ln\Gamma\left(\alpha+\frac12\right) \nonumber \\
& -\frac{1}{2(1-\alpha)} \left\{(1+2\alpha)\ln \alpha + \alpha\ln \frac{\pi}{24} +\ln 6\right\}
\label{renyiMcolorless}
\end{align}
up to terms vanishing as $n\to \infty$. This reproduces (\ref{s1motzkin_renyi1}) with $n_1=n$. 

\subsection*{Colored case ($s>1$)}
As in the Fredkin model, dominant $h$ in the sum (\ref{renyiM}) is different depending on 
the value of $\alpha$. For $0<\alpha<1$  the sum can be evaluated by contribution around the saddle point $h_*=O(n)$, 
whereas for $\alpha>1$ case height $h\leq O(n^0)$ dominantly contributes to the sum.    

The result is as follows (we present the derivation in Appendix~\ref{sec:renyi_Ms}). 
The R\'enyi entropy for \textcolor{red}{$0<\alpha<1$} takes the form
\begin{align}
S_{{\rm M}\,A,\alpha}= &\, n \frac{2\alpha}{1-\alpha}\,\ln\left[\sigma\left(s^{\frac{1-\alpha}{2\alpha}}+s^{-\frac{1-\alpha}{2\alpha}} +s^{-1/2}\right)\right] \nonumber \\
& +\frac{1+\alpha}{2(1-\alpha)}\,\ln n +C(s,\alpha)
\label{renyiMcolor1}
\end{align}
with $C(s,\alpha)$ being $n$-independent terms:
\begin{align} 
C(s,\alpha)\equiv &\, \frac12\ln \pi -\frac{1}{1-\alpha}\ln\left(s\sqrt{\alpha}\right) \nonumber \\
& +\frac{1}{2(1-\alpha)}\,\ln\left(s^{\frac{1}{2\alpha}}+s^{1-\frac{1}{2\alpha}}+4s\right)  \nonumber \\
& +\frac{3\alpha}{2(1-\alpha)}\,\ln(2\sigma)  +\frac{3\alpha-1}{1-\alpha}\,\ln\left(s^{\frac{1}{2\alpha}}+s^{1-\frac{1}{2\alpha}}+1\right) \nonumber \\
&  -\frac{\alpha}{2(1-\alpha)}\,\ln \left[1+4\frac{\left(2s^{\frac{1}{2\alpha}}+1\right)\,\left(2s^{1-\frac{1}{2\alpha}}+1\right)}{\left(s^{\frac{1-\alpha}{2\alpha}}- s^{-\frac{1-\alpha}{2\alpha}} \right)^2}\right] .
\end{align}
The R\'enyi entropy asymptotically grows linearly in $n$, which is similar to the colored Fredkin case. 
The subleading $\ln n$ term seems to have some universal meaning, since its coefficient coincides with that in the colored Fredkin case (\ref{renyiFcolor1}).    
In addition, there is a logarithmic singularity at $\alpha=0$ in $C(s, \alpha)$ of the form $-\frac{1}{2(1-\alpha)}\ln \alpha$ that is in common with colored Fredkin case (\ref{renyiFcolor1}) 
and colorless cases (\ref{s1motzkin_renyi1}), (\ref{s1motzkin_renyi2}), (\ref{renyiFcolorless}), (\ref{renyiMcolorless}). 

For \textcolor{blue}{$\alpha>1$}, we find that the R\'enyi entropy behaves as $\ln n$: 
\begin{align}
S_{{\rm M},\,A,\,\alpha} = & \,\frac{3\alpha}{2(\alpha-1)}\,\ln n -\ln s+\frac{3\alpha}{2(\alpha-1)}\,\ln\sigma \nonumber \\
&+ \frac{\alpha}{2(\alpha-1)}\,\ln \frac{\pi}{2} 
 -\frac{1}{\alpha-1}\,\ln \Phi\left(s^{-(\alpha-1)}, -2\alpha, 0\right) .
\label{renyiMcolor2} 
 \end{align}
The leading term coincides with that in the colored Fredkin case (\ref{renyiFcolor2}) and (\ref{renyiFcolor3}), which 
suggests again that the logarithmic term has some universal meaning.  

In both of the expressions (\ref{renyiMcolor1}) and (\ref{renyiMcolor2}), we cannot take $\alpha\to 1$ or $s\to 1$ limit. 
The limit does not commute with $n\to \infty$ limit. 

\section{Phase Transition}
\setcounter{equation}{0}
In both of colored Fredkin and Motzkin spin chains, we have found that the asymptotic form of the R\'enyi entropy is non-analytic as a function of $\alpha$ at $\alpha=1$. 
This behavior has never seen in any spin chain for which the R\'enyi entropy was computed. 
Even in deformed Fredkin spin chain with $s, t>1$, no such phenomenon is found from~(\ref{dfredkin_renyi}). 

In writing the reduced density matrix in terms of the entanglement Hamiltonian: $\rho_A=e^{-H_{{\rm ent}, \,A}}$, 
$\Tr_A\,\rho_A^\alpha$ is a partition function of this Hamiltonian and the R\'enyi entropy (\ref{renyiA}) is proportional to free energy of the entanglement Hamiltonian. 
The parameter $\alpha$ plays a role of the inverse temperature.   
From this point of view, our result can be interpreted as a phase transition at the inverse temperature $\alpha=1$. 
Colored Dyck/Motzkin paths reaching large height $h=O(n)$ dominantly contribute to the R\'enyi entropy in ``high temperature'' ($0<\alpha<1$), 
whereas paths with low height $h=O(n^0)$ dominate in ``low temperature'' ($\alpha>1$). 
Since the R\'enyi entropy for colorless cases is the qualitatively same as the latter case, highly excited paths of $h=O(n)$ are not activated without colors. 
The transition point $\alpha=1$ itself forms a phase, where the von Neumann entanglement entropy grows as $\sqrt{n}$ and 
paths with height $h= O(\sqrt{n})$ dominate.  
This picture is summarized in Fig.~\ref{fig:phase}.

\section{Discussion}
\setcounter{equation}{0}
We presented a mini review of the R\'enyi entropy in quantum spin chains investigated so far, and 
analytically computed the entropy of a half chain in highly entangled Motzkin and Fredkin spin chains. 
In colored cases, we found non-analyticity in the expression of the R\'enyi entropy at $\alpha=1$, 
which is a totally new phase transition never seen before in spin chains.    
We have been computing the R\'enyi entropy of a block of general length. 
In particular, we conjecture that the same transition at $\alpha=1$ happens for a subsystem $A$ of general length 
including much smaller size than $O(n)$. 
This issue will be reported elsewhere.  
It is an interesting subject to derive the continuum limit of the ground-state wavefunction in the colored cases 
and discuss the phase transition from the viewpoint of continuum field theory as well as in colorless cases~\cite{chen}.  
Finally, it will be intriguing to perform similar computation to cousins of the models~\cite{salberger_etal,zhang_k,zhang_ak,SISmotzkin,SISfredkin,caha_n}.

\section*{Acknowledgement}
F.S. would like to thank the members of Theory Center, KEK, where a part of this work was done and he enjoyed stimulating discussions during his visit. 
V.K. is grateful to Olof Salberger for discussions.

\appendix
\section{The ground state of Fredkin spin chain}
\label{sec:GS_Fredkin}
\setcounter{equation}{0}
The Fredkin spin chain \cite{fredkin} of length $2n$ has up and down quantum spin degrees of freedom with multiplicity (called as color) $s$ at each of the lattice sites $\{1,2,\cdots, 2n\}$. 
We express the up- (down-)spin state with color $k$ at the site $i$ as $\ket{u^k_i}$ ($\ket{d^k_i}$) 
with $k=1,\cdots, s$. 
The Hamiltonian is given by the sum of projection operators: 
\begin{align}
& H_{\rF, s} =  \sum_{j=1}^{2n-2}\sum_{k_1,k_2,k_3=1}^s \left\{\ket{U^{k_1,k_2,k_3}_{j, j+1, j+2}}\bra{U^{k_1,k_2,k_3}_{j, ,j+1, j+2}}+
\ket{D^{k_1,k_2,k_3}_{j, j+1, j+2}}\bra{D^{k_1,k_2,k_3}_{j, j+1, j+2}}\right\} \nn \\
& +\sum_{j=1}^{2n-1} \sum_{k\neq \ell}\left\{\ket{u^k_j, d^\ell_{j+1}}\bra{u^k_j, d^\ell_{j+1}}
+\frac12\left(\ket{u_j^k, d^k_{j+1}}-\ket{u^\ell_j, d^{\ell}_{j+1}}\right)\left(\bra{u_j^k, d^k_{j+1}}-\bra{u^\ell_j, d^{\ell}_{j+1}}\right)\right\} \nn \\
& + \sum_{k=1}^s\left\{\ket{d^k_1}\bra{d^k_1} + \ket{u^k_{2n}}\bra{u^k_{2n}}\right\},
\label{HF}
\end{align}
where
\bea
\ket{U^{k_1,k_2,k_3}_{j, j+1, j+2}} & = & \frac{1}{\sqrt{2}}\left(\ket{u^{k_1}_j, u^{k_2}_{j+1}, d^{k_3}_{j+2}}
-\ket{u^{k_1}_j, d^{k_2}_{j+1}, u^{k_3}_{j+2}}\right), \\
\ket{D^{k_1,k_2,k_3}_{j, j+1, j+2}} & = & \frac{1}{\sqrt{2}}\left(\ket{u^{k_1}_j, d^{k_2}_{j+1}, d^{k_3}_{j+2}}
-\ket{d^{k_1}_j, u^{k_2}_{j+1}, d^{k_3}_{j+2}}\right). 
\eea
The interactions have a local range up to next-to-nearest neighbors. 

For colorless case ($s=1$),  
the up- and down-spin states can be represented as arrows in 2D plane pointing to 
$(1,1)$ (up-step) and $(1,-1)$ (down-step), respectively. 
Each spin configuration of the chain corresponds to a length-$2n$ walk consisting of the up- and down-steps. 
The Hamiltonian has a unique ground state of zero energy, which is superposition of states with equal weight.  
Each of the states is identified with each path of length-$2n$ Dyck walks 
that is random walks starting at the origin, ending at $(2n, 0)$, 
and not allowing paths to enter $y<0$ region.  

\begin{figure}[h!]
\captionsetup{width=0.8\textwidth}
\begin{center}
		\includegraphics[scale=0.8]{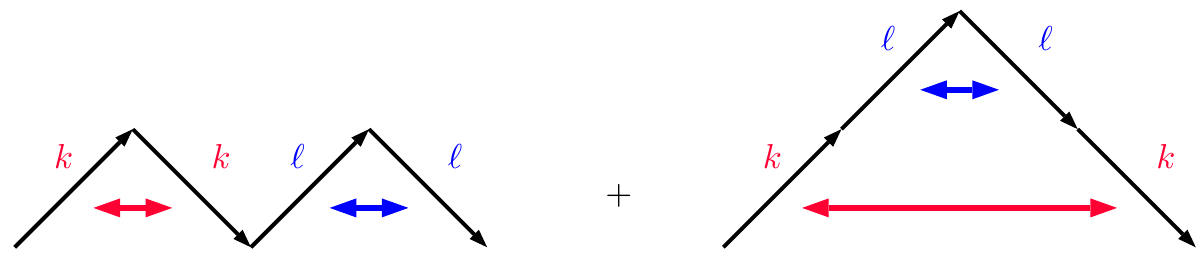} 
	\caption{\small Two terms in the summand of (\ref{PF4}) are depicted. Up- and down-steps with the same color are matched.}
\label{fig:PF4}
\end{center}
\end{figure}

For $s$-color case, 
the above identification goes on with additional color degrees of freedom. 
Namely, each spin configuration of the chain corresponds to a length-$2n$ walk consisting of the up- and 
down-steps with color. 
The ground state is unique, and corresponds to length-$2n$ colored Dyck walks 
in which 
the color of each up-step should be matched with that of the subsequent down-step at the same height. 
The other is the same as the colorless case.
Let $P_{\rF,\, 2n,\,s}$ be the formal sum of length-$2n$ colored Dyck walks. 
For example, $2n=4$ case reads 
\be
P_{\rF,\,4,\,s}= \sum_{k,\ell=1}^s\left\{u^kd^ku^\ell d^\ell + u^k u^\ell d^\ell d^k\right\}, 
\label{PF4}
\ee
where the summand is depicted in Fig.~\ref{fig:PF4}. 
%

The ground state is expressed as 
\be
\ket{P_{\rF,\,2n,\,s}}=\frac{1}{\sqrt{N_{\rF,\,2n,\,s}}}\sum_{w\in P_{\rF,\,2n,\,s}}\ket{w},
\ee
where $w$ runs over monomials appearing in $P_{\rF,\,2n,\,s}$, and 
$N_{\rF,\,2n,\,s}$ stands for the number of the length-$2n$ colored Dyck walks: 
\be
N_{\rF,\,2n,\,s}=s^n\,N_{\rF,\,2n}=\frac{s^n}{n+1}\binomi{2n}{n}.
\label{N2n_F}
\ee
$N_{\rF,\,2n}$ in the middle denotes the number 
of colorless Dyck walks of length $2n$, which is equal to the Catalan number $C_n$. 
Note that $N_{\rF,\,2n,\,s}$ can be obtained by setting all the $u^k$ and $d^k$ to 1 in $P_{\rF,\,2n,\,s}$. 
The $2n=4$ case reads  
\be
\ket{P_{\rF,\, 4,\, s}}=\frac{1}{\sqrt{2s^2}}\sum_{k,\ell=1}^s
\left\{\ket{u^k_1, d^k_2, u^\ell_3, d^\ell_4} + \ket{u^k_1, u^\ell_2, d^\ell_3,d^k_4}\right\}.
\ee

We divide the full system $S=\{1, 2, \cdots, 2n\}$ 
into two subsystems $A=\{1,2, \cdots, n\}$ and $B=\{ n+1, n+2, \cdots, 2n\}$ of the same size, and 
compute the R\'enyi entropy of $A$: 
\be
S_{A, \alpha}\equiv \frac{1}{1-\alpha}\,\ln \Tr_A\,\rho_A^\alpha, 
\label{renyi}
\ee
where $\rho_A=\Tr_B\, \rho$ is the reduced density matrix of $A$ obtained by tracing out the Hilbert space belonging to $B$. 
The parameter $\alpha$ is positive, and not equal to 1. 
The limit $\alpha\to 1$ reproduces the von Neumann entanglement entropy
\be
S_A=-\Tr_A\left(\rho_A\ln\rho_A\right) . 
\label{vonneumann}
\ee

\begin{figure}[h!]
\captionsetup{width=0.8\textwidth}
\begin{center}
		\includegraphics[scale=0.8]{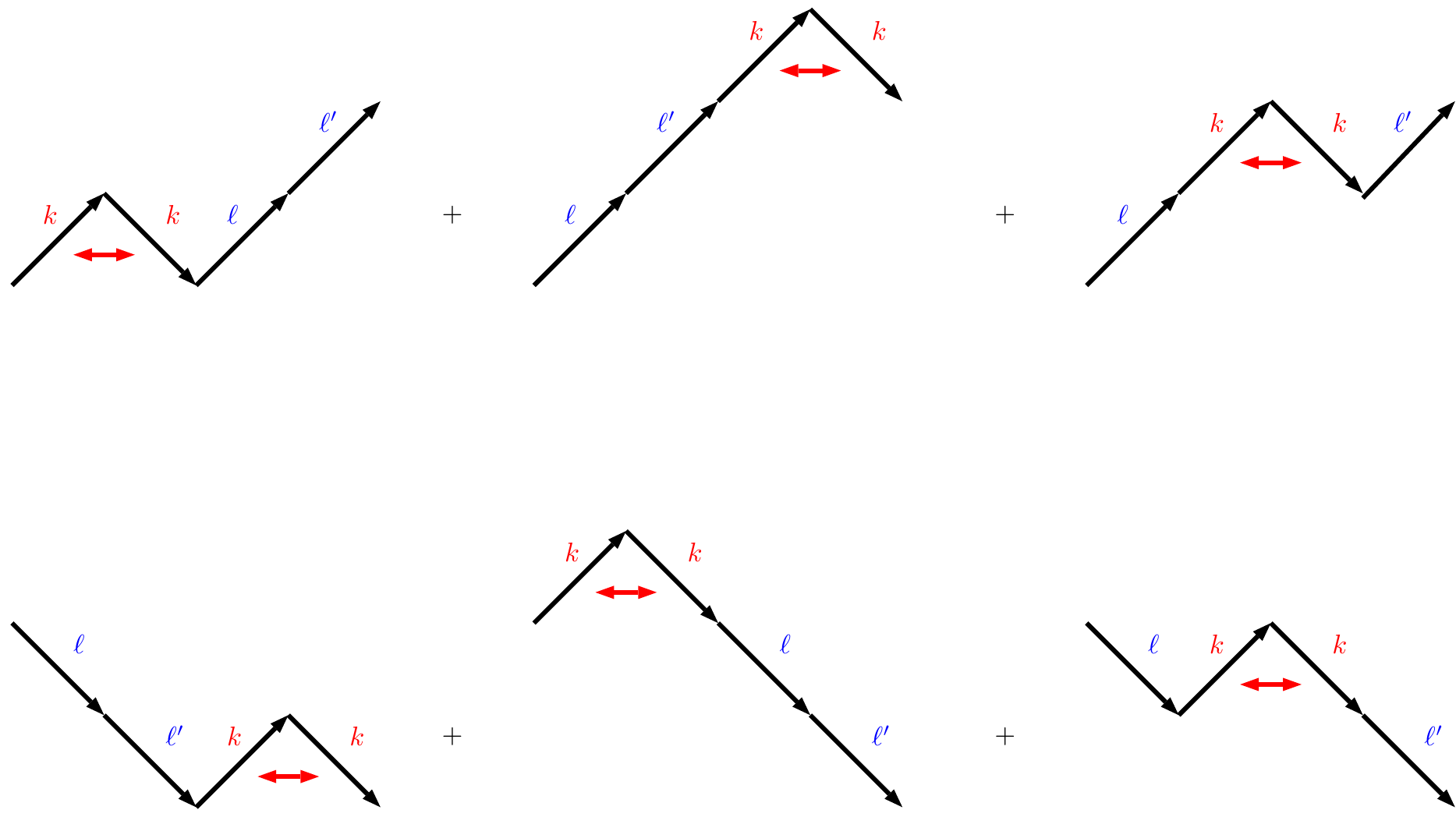} 
	\caption{\small Three terms in the summand of (\ref{P4h02_F}) are depicted in the first line.
	 Each term contains two unmatched up-steps. 
	 Also, three terms in the summand of (\ref{P4h20_F}) are illustrated in the second line. Each has two unmatched down-steps.}
\label{fig:PF4h2}
\end{center}
\end{figure}

Let us consider a part of colored Dyck paths belonging to $A$, which consists of paths from $(0,0)$ to $(n, h)$ 
denoted by $P_{\rF,\, n, \,s}^{(0\to h)}$. Height $h$ takes nonnegative integers. 
Similarly, the remaining part of paths belonging to $B$ is denoted by $P_{\rF,\, n, \,s}^{(h\to 0)}$. 
Each path in $P_{\rF,\, n, \,s}^{(0\to h)}$ has $h$ unmatched up-steps that are supposed to be 
matched with $h$ down-steps in $P_{\rF,\, n, \,s}^{(h\to 0)}$. 
Let $\tilde{P}_{\rF,\, n,\,s}^{(0\to h)}(\{\kappa_m\})$ ($\tilde{P}_{\rF,\, n, \,s}^{(h\to 0)}(\{\kappa_m\})$) 
be paths obtained by freezing the color degrees of freedom of 
$h$ unmatched up-steps in $P_{\rF,\, n, \,s}^{(0\to h)}$ (down-steps in $P_{\rF,\, n, \,s}^{(h\to 0)}$), 
where up-  and down-steps with frozen color between heights $m-1$ and $m$ are expressed as $u^{\kappa_m}$ and $d^{\kappa_m}$.  
As an example, case of $2n=4$ and $h=2$ gives 
\begin{align}
& P^{(0\to 2)}_{\rF,\,4,\,s} = \sum_{k,\ell, \ell'=1}^s\left\{u^k d^ku^\ell u^{\ell'} + u^\ell u^{\ell'} u^k d^k + u^\ell u^k d^k u^{\ell'}\right\}, 
\label{P4h02_F}\\
& P^{(2\to 0)}_{\rF,\,4,\,s} = \sum_{k,\ell, \ell'=1}^s\left\{d^\ell d^{\ell'} u^k d^k + u^k d^k d^\ell d^{\ell'} + d^\ell u^k d^k d^{\ell'}\right\}, 
\label{P4h20_F}\\
& \tilde{P}^{(0\to 2)}_{\rF,\,4,\,s}(\{\kappa_m\}) = \left(\sum_{k=1}^s u^k d^k\right) u^{\kappa_1}u^{\kappa_2} 
+u^{\kappa_1}u^{\kappa_2} \left(\sum_{k=1}^s u^k d^k\right) + u^{\kappa_1}\left(\sum_{k=1}^s u^k d^k\right) u^{\kappa_2}, \\
& \tilde{P}^{(2\to 0)}_{\rF,\,4,\,s}(\{\kappa_m\}) = d^{\kappa_2} d^{\kappa_1} \left(\sum_{k=1}^s u^k d^k\right) 
+ \left(\sum_{k=1}^s u^k d^k\right)d^{\kappa_2} d^{\kappa_1}  + d^{\kappa_2}\left(\sum_{k=1}^s u^k d^k\right) d^{\kappa_1} ,
\end{align}
where $u^\ell$, $u^{\ell'}$, $d^\ell$ and $d^{\ell'}$ are unmatched up- and down-steps in (\ref{P4h02_F}) and (\ref{P4h20_F}). 
The summands of (\ref{P4h02_F}) and (\ref{P4h20_F}) are depicted in Fig.~\ref{fig:PF4h2}. 
%

The numbers of the paths are given by 
\be
N_{\rF,\, n,\, s}^{(0\to h)} = \left.P_{\rF,\, n, \,s}^{(0\to h)}\right|_{u^k=d^k=1} = s^{\frac{n+h}{2}}N^{(h)}_{\rF,\,n},
\qquad 
\tilde{N}_{\rF,\,n,\,s}^{(0\to h)} = s^{-h} N_{\rF,\, n,\, s}^{(0\to h)} =s^{\frac{n-h}{2}}N^{(h)}_{\rF,\,n}
\label{Ntilde_F}
\ee
with 
\be
N^{(h)}_{\rF,\,n}=\frac{1+(-1)^{n+h}}{2}\frac{h+1}{\frac{n+h}{2}+1}\binomi{n}{\frac{n+h}{2}}.
\label{Nhn_F}
\ee
By considering the reverse of paths, it is easy to see $N_{\rF,\, n,\, s}^{(0\to h)} =N_{\rF,\, n,\, s}^{(h\to 0)}$ and 
$\tilde{N}_{\rF,\,n,\,s}^{(0\to h)}=\tilde{N}_{\rF,\,n,\,s}^{(h\to 0)}$.   
Then, we can decompose $P_{\rF,\, 2n,\,s}$ into the two halves of chains as 
\be
P_{\rF,\,2n,\,s}=\sum_{h=0}^n\sum_{\kappa_1=1}^s\cdots \sum_{\kappa_h=1}^s 
\tilde{P}^{(0\to h)}_{\rF,\, n,\,s}(\{\kappa_m\})\,\tilde{P}^{(h\to 0)}_{\rF,\, n,\,s}(\{\kappa_m\}).
\label{Schmidt_F0}
\ee
The sums of $\kappa_1,\cdots, \kappa_h$ provide the match of colors for up-steps in $A$ and down-steps in $B$. 
This leads to the Schmidt decomposition:
\be
\ket{P_{\rF,\,2n,\,s}} = \sum_{h=0}^n\sum_{\kappa_1=1}^s\cdots \sum_{\kappa_h=1}^s \sqrt{p^{(h)}_{\rF,\,n,n,\,s}}\,
\ket{\tilde{P}^{(0\to h)}_{\rF,\, n,\,s}(\{\kappa_m\})}\otimes\ket{\tilde{P}^{(h\to 0)}_{\rF,\, n,\,s}(\{\kappa_m\})}. 
\label{Schmidt_F}
\ee
Here, the states 
$\ket{\tilde{P}^{(0\to h)}_{\rF,\, n,\,s}(\{\kappa_m\})}$ and $\ket{\tilde{P}^{(h\to 0)}_{\rF,\, n,\,s}(\{\kappa_m\})}$ are normalized as 
\bea
\ket{\tilde{P}^{(0\to h)}_{\rF,\, n,\,s}(\{\kappa_m\})} & = & 
\frac{1}{\sqrt{\tilde{N}_{\rF,\,n,\,s}^{(0\to h)}}}\,\sum_{w\in\tilde{P}^{(0\to h)}_{\rF,\, n,\,s}(\{\kappa_m\})}\ket{w}, \\
\ket{\tilde{P}^{(h\to 0)}_{\rF,\, n,\,s}(\{\kappa_m\})} & = &
\frac{1}{\sqrt{\tilde{N}_{\rF,\,n,\,s}^{(0\to h)}}}\,\sum_{w\in\tilde{P}^{(h\to 0)}_{\rF,\, n,\,s}(\{\kappa_m\})}\ket{w}, 
\eea
and 
\be
p^{(h)}_{\rF,\,n,n,\,s}=\frac{\left(\tilde{N}_{\rF,\,n,\,s}^{(0\to h)}\right)^2}{N_{\rF,\, 2n, \,s}}
=s^{-h}\frac{\left(N^{(h)}_{\rF,\,n}\right)^2}{N_{\rF,\, 2n}}.
\label{pnn_F}
\ee

From the density matrix of the ground state $\rho=\ket{P_{\rF,\,2n,\,s}}\bra{P_{\rF,\,2n,\,s}}$ with (\ref{Schmidt_F}), 
the reduced density matrix is obtained as 
\be
\rho_A = \Tr_B\,\rho= \sum_{h=0}^n\sum_{\kappa_1=1}^s\cdots \sum_{\kappa_h=1}^s p^{(h)}_{\rF,\,n,n,\,s}\,
\ket{\tilde{P}^{(0\to h)}_{\rF,\, n,\,s}(\{\kappa_m\})}\bra{\tilde{P}^{(0\to h)}_{\rF,\, n,\,s}(\{\kappa_m\})}, 
\label{rhoA_F}
\ee
where we used the orthonormal property: 
\be
\left\langle\tilde{P}^{(h\to 0)}_{\rF,\, n,\,s}(\{\kappa_m\})\right.\ket{\tilde{P}^{(h'\to 0)}_{\rF,\, n,\,s}(\{\kappa'_m\})} 
=\delta_{h,h'}\delta_{\kappa_1,\kappa'_1}\cdots \delta_{\kappa_h,\kappa'_h}.
\ee
Since $\rho_A$ is a diagonal form, the R\'enyi entropy (\ref{renyi}) becomes 
\be
S_{\rF,\,A,\alpha} = \frac{1}{1-\alpha}\,\ln \sum_{h\geq 0}s^h\,\left(p^{(h)}_{\rF,\, n,n,\,s}\right)^\alpha. 
\label{renyi_F}
\ee 
Note $p^{(h)}_{\rF,\, n,n,\,s}$ does not depend on $\kappa_1, \cdots, \kappa_h$ 
and the sums $\sum_{\kappa_1=1}^s\cdots \sum_{\kappa_h=1}^s$ yield the factor $s^h$. 

\paragraph{Asymptotic form of $p^{(h)}_{\rF, \, n,n,\,s}$:}
We plug (\ref{N2n_F}) and (\ref{Nhn_F}) to (\ref{pnn_F}) and evaluate its asymptotic behavior. 
For $n, \,n\pm h\gg 1$, use of Stirling's formula $x!\simeq \sqrt{2\pi}\,x^{x+\frac12}\,e^{-x}\,\left[1+O(x^{-1})\right]$ ($x\to \infty$) 
leads to
\bea
p^{(h)}_{\rF,\,n,n,\,s} & \simeq & 
\frac{1+(-1)^{n+h}}{2}\,\frac{8}{\sqrt{\pi}}\,s^{-h}\frac{(h+1)^2}{n^{3/2}}\nn \\
& & \times \exp\left[-(n+h+3)\ln \frac{n+h}{n}-(n-h+1)\ln\frac{n-h}{n} +O(n^{-1})\right]. 
\label{pnn_asym_F}
\eea
In case of $h\leq O(n^{1/2})$, we further expand the logarithms in the exponential in (\ref{pnn_asym_F}) to obtain 
\be
 p^{(h)}_{\rF,\,n,n,\,s} \simeq 
\frac{1+(-1)^{n+h}}{2}\,\frac{8}{\sqrt{\pi}}\,s^{-h}\frac{(h+1)^2}{n^{3/2}}\,e^{-\frac{(h+1)^2}{n}}\times \left[1+O(n^{-1})\right].
\label{pnn_asym_F2}
\ee

\section{The ground state of Motzkin spin chain} 
\label{sec:GS_Motzkin}
\setcounter{equation}{0}
The Motzkin spin chain \cite{motzkin} has additional spin degrees of freedom (we call zero-spin) at each site compared with the Fredkin spin chain. 
We express the up- and down-spin states with color $k=1,\cdots, s$ and the zero-spin at the site $i$ as $\ket{u^k_i}$, 
$\ket{d^k_i}$ and $\ket{0_i}$, respectively. 
The Hamiltonian of the Motzkin spin chain of length $2n$ is 
\bea
H_{\rM, s} & = & \sum_{j=1}^{2n-1}\sum_{k=1}^s \left\{\ket{U^k_{j, j+1}}\bra{U^k_{j, j+1}}+
\ket{D^k_{j, j+1}}\bra{D^k_{j, j+1}} + \ket{F^k_{j, j+1}}\bra{F^k_{j, j+1}}\right\} \nn \\
& & +\sum_{j=1}^{2n-1}\sum_{k\neq\ell}\ket{u^k_j, d^\ell_{j+1}}\bra{u^k_j, d^\ell_{j+1}} 
+ \sum_{k=1}^s\left\{\ket{d^k_1}\bra{d^k_1} + \ket{u^k_{2n}}\bra{u^k_{2n}}\right\},
\label{HM}
\eea
where
\bea
\ket{U^k_{j, j+1}} & = & \frac{1}{\sqrt{2}}\left(\ket{0_j, u^k_{j+1}}
-\ket{u^k_j, 0_{j+1}}\right), \\
\ket{D^k_{j, j+1}} & = & \frac{1}{\sqrt{2}}\left(\ket{0_j, d^k_{j+1}}
-\ket{d^k_j, 0_{j+1}}\right),\\
\ket{F^k_{j, j+1}} & = & \frac{1}{\sqrt{2}}\left(\ket{0_j, 0_{j+1}}
-\ket{u^k_j, d^k_{j+1}}\right),
\eea
and the interactions are among nearest neighbors. 

The Hamiltonian has the unique ground state of zero-energy. 
By the same identification of the spins and 2D steps as before with the zero-spin corresponding to the arrow $(1,0)$ (flat-step), 
for colorless case ($s=1$) the ground state is expressed by the equal-weight superposition of length-$2n$ Motzkin walks, 
which are random walks consisting of up-, down- and flat-steps, 
starting at the origin, ending at $(2n, 0)$ and not allowing paths to enter $y<0$ region. 
For $s$-colored case, the color assigned to each up-step 
should be matched with that of the subsequent down-step at the same height. 

The ground state is expressed as 
\be
\ket{P_{\rM,\,2n,\,s}}=\frac{1}{\sqrt{N_{\rM,\,2n,\,s}}}\sum_{w\in P_{\rM,\,2n,\,s}}\ket{w},
\ee
where $N_{\rM,\,2n,\,s}$ in the normalization factor is the number of the length-$2n$ colored Motzkin walks given 
by 
\be
N_{\rM,\,2n,\,s}=\sum_{\rho=0}^{n}\binomi{2n}{2\rho}s^{n-\rho}\,N_{\rF,\,2n-2\rho},
\label{N2n_M}
\ee
where $2\rho$ stands for the number of the flat-steps. 
For example, 
\begin{align}
\ket{P_{\rM,\, 4,\, s}} = &\,\frac{1}{\sqrt{1+6s+2s^2}}\Biggl[  \ket{0_1, 0_2, 0_3, 0_4}  \nn \\
& \hspace{20mm}+ \sum_{k=1}^s \left\{\ket{u^k_1, d^k_2, 0_3, 0_4} + \ket{0_1, u^k_2, d^k_3, 0_4} +\ket{0_1, 0_2, u^k_3, d^k_4} \right.\nn \\
& \hspace{30mm}\left.+\ket{u^k_1, 0_2, d^k_3, 0_4} + \ket{0_1, u^k_2, 0_3, d^k_4} + \ket{u^k_1, 0_2, 0_3, d^k_4}\right\} \nn \\
& \hspace{20mm}+\sum_{k,\ell=1}^s \left\{\ket{u^k_1, d^k_2, u^\ell_3, d^\ell_4} + \ket{u^k_1, u^\ell_2, d^\ell_3,d^k_4}\right\}\Biggr].
\end{align}

\begin{figure}[h!]
\captionsetup{width=0.8\textwidth}
\begin{center}
		\includegraphics[scale=0.8]{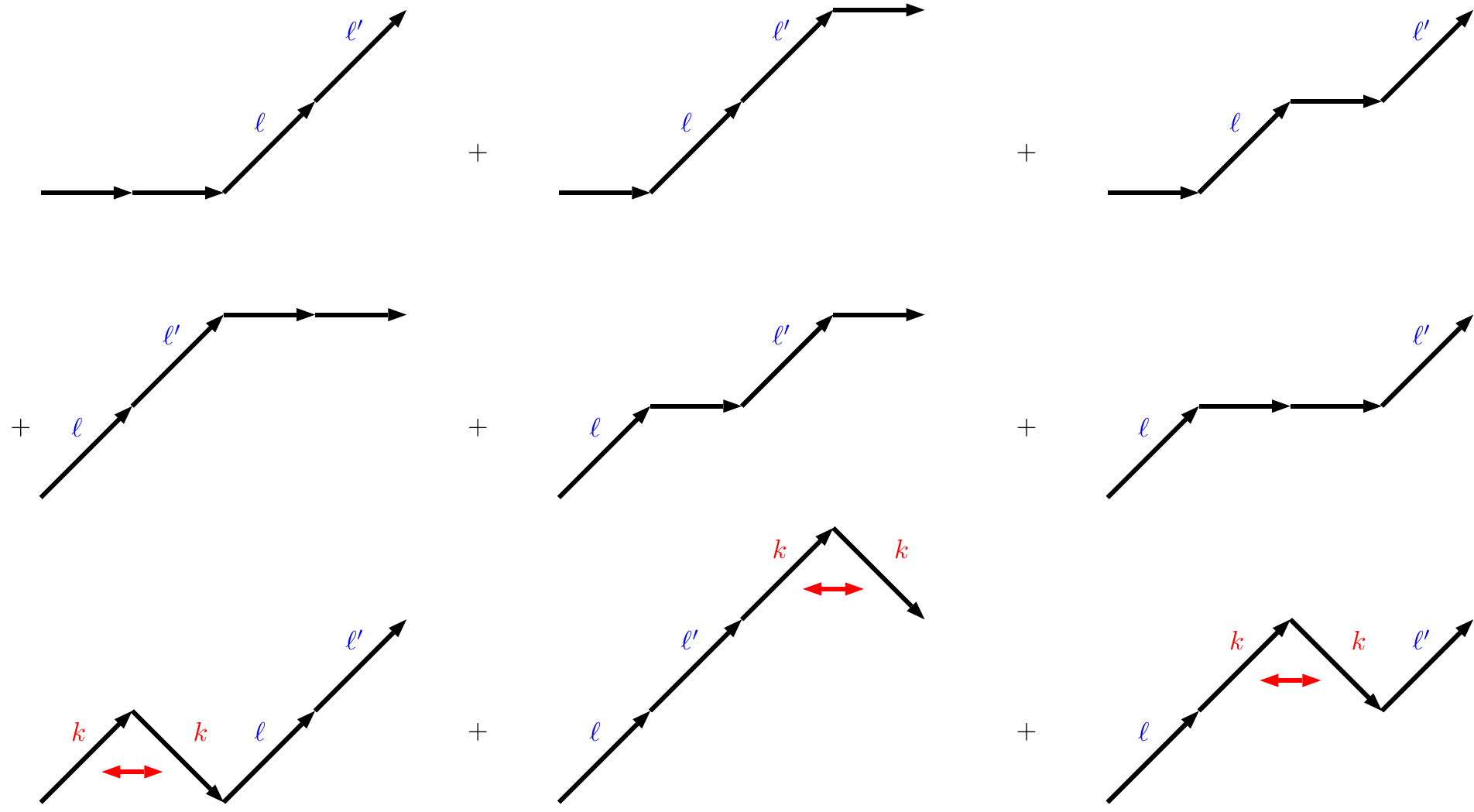} 
	\caption{\small Six terms in the summand of the first sum in (\ref{P4h02_M}) are depicted in the first two lines.
	 Three terms in the summand of the second sum in (\ref{P4h02_M}) are illustrated in the last line.}
\label{fig:PM4h02}
\end{center}
\end{figure}

\begin{figure}[h!]
\captionsetup{width=0.8\textwidth}
\begin{center}
		\includegraphics[scale=0.8]{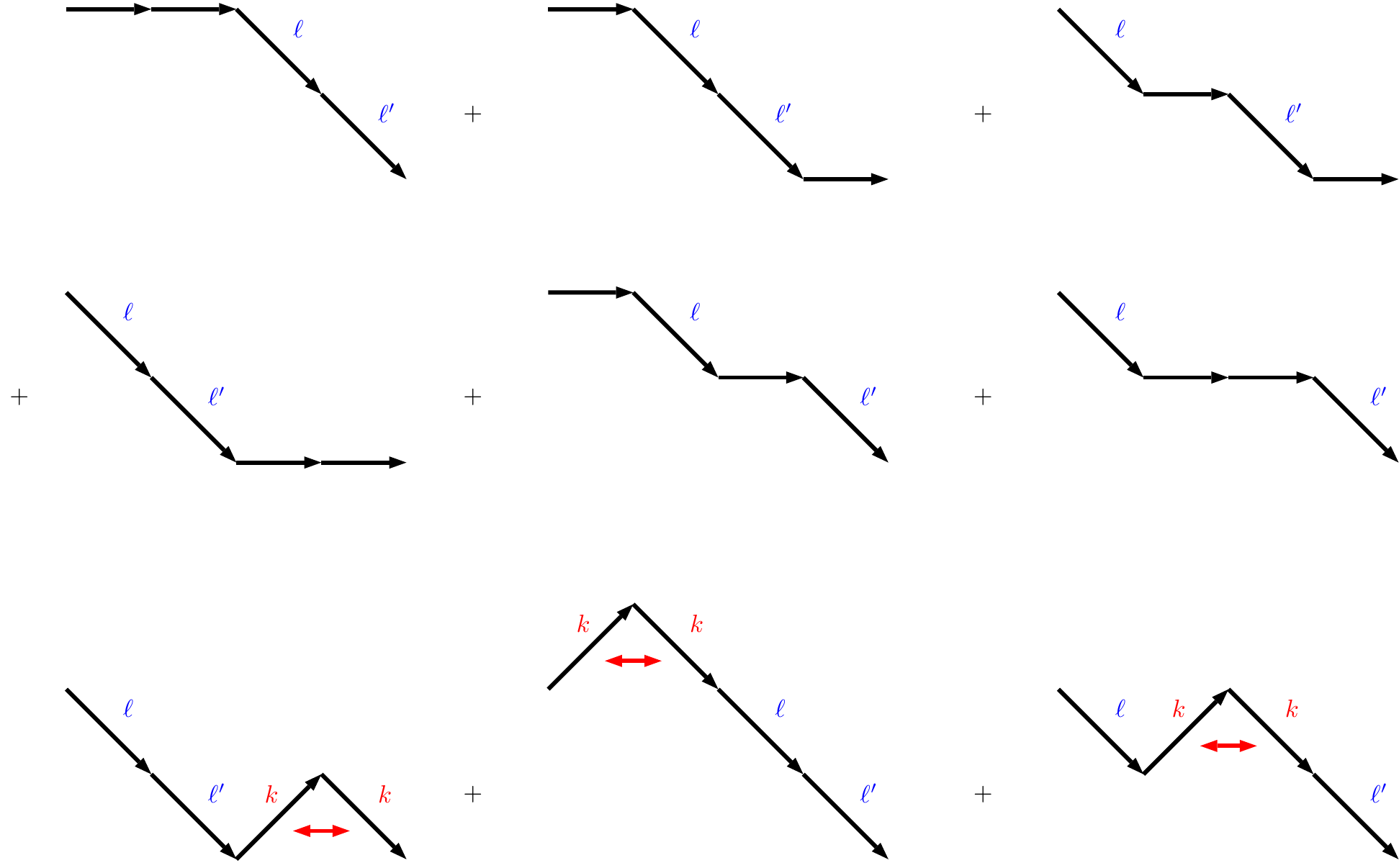} 
	\caption{\small Six terms in the summand of the first sum in (\ref{P4h20_M}) are depicted in the first two lines.
	 Three terms in the summand of the second sum in (\ref{P4h20_M}) are illustrated in the last line.}
\label{fig:PM4h20}
\end{center}
\end{figure}

We divide the system into the two subsystems $A$ and $B$ same as in the Fredkin case. 
$P_{\rM,\, n, \,s}^{(0\to h)}$ denotes paths from $(0,0)$ to $(n, h)$ which are a part of colored Motzkin paths belonging to $A$.  
Similarly, $P_{\rM,\, n, \,s}^{(h\to 0)}$ stands for paths from $(n,h)$ to $(0,0)$ which belong to $B$. 
Each path in $P_{\rM,\, n, \,s}^{(0\to h)}$ contains $h$ unmatched up-steps that should be 
matched with $h$ down-steps in $P_{\rM,\, n, \,s}^{(h\to 0)}$. 
Let $\tilde{P}_{\rM,\, n,\,s}^{(0\to h)}(\{\kappa_m\})$ ($\tilde{P}_{\rM,\, n, \,s}^{(h\to 0)}(\{\kappa_m\})$) 
be paths obtained by freezing the color degrees of freedom of 
$h$ unmatched up-steps in $P_{\rM,\, n, \,s}^{(0\to h)}$ to $u^{\kappa_1},\cdots, u^{\kappa_h}$ 
(down-steps in $P_{\rM,\, n, \,s}^{(h\to 0)}$ to $d^{\kappa_1},\cdots, d^{\kappa_h}$). 
$u^{\kappa_m}$ and $d^{\kappa_m}$ stand for up- and down-steps between the heights $m-1$ and $m$. 
For example, $2n=4$ and $h=2$ case gives
\bea
P^{(0\to 2)}_{\rM,\,4,\,s} & = & \sum_{\ell,\ell'=1}^s \left\{00u^\ell u^{\ell'} + 0u^\ell u^{\ell'}0 + 0u^\ell 0 u^{\ell'} 
+ u^\ell u^{\ell'} 00 + u^\ell 0u^{\ell'} 0 + u^\ell 00u^{\ell'}\right\} \nn \\
& & +\sum_{k,\ell, \ell'=1}^s\left\{u^k d^ku^\ell u^{\ell'} + u^\ell u^{\ell'} u^k d^k + u^\ell u^k d^k u^{\ell'}\right\}, 
\label{P4h02_M}\\
P^{(2\to 0)}_{\rM,\,4,\,s} & = & \sum_{\ell,\ell'=1}^s \left\{00d^\ell d^{\ell'} + 0d^\ell d^{\ell'}0 + d^\ell 0 d^{\ell'} 0 
+ d^\ell d^{\ell'}00 + 0d^\ell 0d^{\ell'} + d^\ell 00d^{\ell'}\right\} \nn \\
& & +\sum_{k,\ell, \ell'=1}^s\left\{d^\ell d^{\ell'} u^k d^k + u^k d^k d^\ell d^{\ell'} + d^\ell u^k d^k d^{\ell'}\right\}, 
\label{P4h20_M}
\\
\tilde{P}^{(0\to 2)}_{\rM,\,4,\,s}(\{\kappa_m\}) & = & 00u^{\kappa_1}u^{\kappa_2}  + 0u^{\kappa_1}u^{\kappa_2} 0 + 0u^{\kappa_1}0u^{\kappa_2} 
+u^{\kappa_1}u^{\kappa_2} 00 + u^{\kappa_1}0u^{\kappa_2}0 + u^{\kappa_1}00u^{\kappa_2}  \nn \\
& & \hspace{-7mm}+\left(\sum_{k=1}^s u^k d^k\right) u^{\kappa_1}u^{\kappa_2} 
+u^{\kappa_1}u^{\kappa_2} \left(\sum_{k=1}^s u^k d^k\right) + u^{\kappa_1}\left(\sum_{k=1}^s u^k d^k\right) u^{\kappa_2}, \\
\tilde{P}^{(2\to 0)}_{\rM,\,4,\,s}(\{\kappa_m\}) & = & 00d^{\kappa_2}d^{\kappa_1}  + 0d^{\kappa_2}d^{\kappa_1} 0 + d^{\kappa_2}d^{\kappa_1} 0
+00d^{\kappa_2}d^{\kappa_1} + 0 d^{\kappa_2}0d^{\kappa_1} + d^{\kappa_2}00d^{\kappa_1}  \nn \\
& & \hspace{-7mm}+d^{\kappa_2} d^{\kappa_1} \left(\sum_{k=1}^s u^k d^k\right) 
+ \left(\sum_{k=1}^s u^k d^k\right) d^{\kappa_2} d^{\kappa_1}  + d^{\kappa_2}\left(\sum_{k=1}^s u^k d^k\right) d^{\kappa_1}.
\eea
The summands of (\ref{P4h02_M}) and (\ref{P4h20_M}) are depicted in Figs.~\ref{fig:PM4h02} and \ref{fig:PM4h20}. 
%
%

The numbers of the paths are 
\be
N_{\rM,\, n,\, s}^{(0\to h)} = s^h\,\tilde{N}_{\rM,\,n,\,s}^{(0\to h)} 
=\sum_{\rho=0}^{n-h}\binomi{n}{\rho} N^{(h)}_{\rF,\, n-\rho}\,s^{\frac{n-\rho+h}{2}}.
\label{Ntilde_M}
\ee
where $N^{(h)}_{\rF,\,n}$ is given in (\ref{Nhn_F}). 
$N_{\rM,\, n,\, s}^{(0\to h)} =N_{\rM,\, n,\, s}^{(h\to 0)}$ and 
$\tilde{N}_{\rM,\,n,\,s}^{(0\to h)}=\tilde{N}_{\rM,\,n,\,s}^{(h\to 0)}$ hold.   
We find a similar decomposition to the Fredkin case:
\be
\ket{P_{\rM,\,2n,\,s}}=\sum_{h=0}^n\sum_{\kappa_1=1}^s\cdots 
\sum_{\kappa_h=1}^s \sqrt{p^{(h)}_{\rM,\,n,n,\,s}}\,
\ket{\tilde{P}^{(0\to h)}_{\rM,\, n,\,s}(\{\kappa_m\})}\otimes\ket{\tilde{P}^{(h\to 0)}_{\rM,\, n,\,s}(\{\kappa_m\})}, 
\label{Schmidt_M}
\ee
where 
\bea
\ket{\tilde{P}^{(0\to h)}_{\rM,\, n,\,s}(\{\kappa_m\})} & = & \frac{1}{\sqrt{\tilde{N}_{\rM,\,n,\,s}^{(0\to h)}}}\,
\sum_{w\in\tilde{P}^{(0\to h)}_{\rM,\, n,\,s}(\{\kappa_m\})}\ket{w}, \\
\ket{\tilde{P}^{(h\to 0)}_{\rM,\, n,\,s}(\{\kappa_m\})} & = & \frac{1}{\sqrt{\tilde{N}_{\rM,\,n,\,s}^{(0\to h)}}}\,
\sum_{w\in\tilde{P}^{(h\to 0)}_{\rM,\, n,\,s}(\{\kappa_m\})}\ket{w}, 
\eea 
and 
\be
p^{(h)}_{\rM,\,n,n,\,s}=\frac{\left(\tilde{N}_{\rM,\,n,\,s}^{(0\to h)}\right)^2}{N_{\rM,\, 2n, \,s}}.
\label{pnn_M}
\ee
This provides the reduced density matrix and the R\'enyi entropy as 
\bea
& & \rho_A = \Tr_B\,\rho= \sum_{h=0}^n\sum_{\kappa_1=1}^s\cdots \sum_{\kappa_h=1}^s p^{(h)}_{\rM,\,n,n,\,s}\,
\ket{\tilde{P}^{(0\to h)}_{\rM,\, n,\,s}(\{\kappa_m\})}\bra{\tilde{P}^{(0\to h)}_{\rM,\, n,\,s}(\{\kappa_m\})}, 
\label{rhoA_M}
\\
 & & S_{\rM,\,A,\alpha} = \frac{1}{1-\alpha}\,\ln \sum_{h= 0}^n s^h\,\left(p^{(h)}_{\rM,\, n,n,\,s}\right)^\alpha. 
\label{renyi_M}
\eea

\paragraph{Asymptotic form of $p^{(h)}_{\rM, \, n,n,\,s}$:}
We rewrite (\ref{Ntilde_M}) as 
\be
\tilde{N}_{\rM,\,n,\,s}^{(0\to h)}=(h+1)\sum_{\rho=0}^{n-h}\frac{1+(-1)^{n-\rho+h}}{2} \,C_{n,h,\rho}
\label{Ntilde_M2}
\ee
with
\be
C_{n,h,\rho}\equiv \frac{n!\,s^{\frac{n-\rho-h}{2}}}{\rho!\left(\frac{n-\rho-h}{2}\right)!\left(\frac{n-\rho+h}{2}+1\right)!}.
\label{Cnhr}
\ee
By using Stirling's formula for $n, \, \rho,\, n-\rho\pm h\gg1$, we obtain 
\bea
C_{n,h,\rho} & \simeq & \frac{2^{n-\rho+1}}{\pi}\frac{n^{n+\frac12}s^{\frac{n-\rho-h}{2}}}{\rho^{\rho+\frac12}(n-\rho)^{n-\rho+2}} \nn \\
& & \times \exp\left[-\frac{n-\rho-h+1}{2}\ln \frac{n-\rho-h}{n-\rho}  -\frac{n-\rho+h+3}{2}\ln \frac{n-\rho+h}{n-\rho} \right] \nn \\
& & \times \left[1+ O(n^{-1})\right].
\label{Cnhr2}
\eea

We evaluate the sum of $\rho$ in (\ref{Ntilde_M2}) by the saddle point method. 
The saddle point equation 
\be 
C_{n,h,\rho}=C_{n,h,\rho+2}, 
\ee
leads to 
\be
(4s-1)\xi^2+2\xi +\eta^2 -1 = 0 +O(n^{-1}) 
\ee
with $\xi=\rho/n$ and $\eta=h/n$. 
The solution is given by 
\be
\rho=\rho_0+O(n^0), \qquad \rho_0\equiv \frac{n}{4s-1}\left[-1+\sqrt{4s-(4s-1)\frac{h^2}{n^2}}\right]. 
\label{saddlepoint_r}
\ee
Here, we take the ``$+$" branch, since $\rho_0$ should be positive.  
After expanding (\ref{Cnhr2}) around the saddle point as $\rho=\rho_0+x$ ($x\ll n$), we have~\footnote{
The relation $(n-\rho_0)^2-h^2=4s\rho_0^2$ is useful to obtain (\ref{Cnhr3}). 
Since $\rho_0$ is an approximate saddle point ($O(n^0)$ terms in (\ref{saddlepoint_r}) are dropped), the expansion has a linear term in $x$.
} 
\bea
C_{n,h,\rho} & \simeq & \frac{s^{-h/2}}{2\pi s}\,\frac{n^{n+\frac12}}{\rho_0^{n+\frac52}} 
\left(\frac{n-\rho_0-h}{n-\rho_0+h}\right)^{\frac{h+1}{2}} 
\exp\left[-\left\{\frac{1}{8s\rho_0^2}\sqrt{4sn^2-(4s-1)h^2} +O(n^{-2})\right\} \,x^2 \right] \nn \\
& & \times \exp\left[\frac{1}{4s\rho_0^2}\left(2n-2(n+1)\rho_0-h\right)x +O\left(x^3n^{-2}, x^4n^{-3},\cdots\right)\right]. 
\label{Cnhr3}
\eea
The sum in (\ref{Ntilde_M2}) can be carried out by converting an integral as 
\be
\sum_{\rho=0}^{n-h} \frac{1+(-1)^{n-\rho+h}}{2}\,(\cdots)  \to \frac12\int_{-\infty}^{\infty} dx \,(\cdots),
\ee 
where we perform Gaussian integrations after expanding the exponential in the last line of (\ref{Cnhr3}). 
Note that $x$ can be regarded as at most $O(\sqrt{n})$ in the Gaussian integral. In addition, parity-odd terms 
with respect to $x\to -x$ vanish in the integral. 
As a result, contribution from the last line amounts to a factor $\left[1+ O(n^{-1})\right]$. 
We finally obtain 
\be
\tilde{N}_{\rM,\,n,\,s}^{(0\to h)} \simeq  
\frac{s^{-h/2}}{\sqrt{2\pi s}}\,\frac{n^{n+\frac12}}{\rho_0^{n+\frac32}}\,\frac{h+1}{\left[4sn^2-(4s-1)h^2\right]^{1/4}} 
\left(\frac{n-\rho_0-h}{n-\rho_0+h}\right)^{\frac{h+1}{2}} \, \times \left[1+ O(n^{-1})\right].
\label{Ntilde_asym_M}
\ee
Replacing $n$ by $2n$ and setting $h=0$ in (\ref{Ntilde_asym_M}) gives the asymptotic form of (\ref{N2n_M}):
\be
N_{\rM,\,2n,\,s}\simeq \frac{\left(2\sqrt{s}+1\right)^{2n+\frac32}}{2\sqrt{\pi}\,s^{3/4}(2n)^{3/2}}\,\times \left[1+ O(n^{-1})\right].
\label{N2n_asym_M}
\ee
Plugging (\ref{Ntilde_asym_M}) and (\ref{N2n_asym_M}) to (\ref{pnn_M}), we find 
\bea
p^{(h)}_{\rM, \, n,n,\,s} & \simeq & 
\frac{s^{-h}}{\sqrt{\pi}\,s^{1/4}}\,\frac{(2n)^{3/2}}{\left(2\sqrt{s}+1\right)^{2n+\frac32}}\,\frac{n^{2n+1}}{\rho_0^{2n+3}} \,
\frac{(h+1)^2}{\left[4sn^2-(4s-1)h^2\right]^{1/2}}\,\left(\frac{n-\rho_0-h}{n-\rho_0+h}\right)^{h+1} \nn \\
& & \times \left[1+ O(n^{-1})\right].
\label{pnn_asym_M}
\eea

When $h$ is at most $O(n^{1/2})$, we have 
\bea
\rho_0 & = & \frac{n}{2\sqrt{s}+1}-\frac{1}{4\sqrt{s}}\frac{h^2}{n} +O(n^{-1}), \\
\left(\frac{n}{\rho_0}\right)^{2n} & = & \left(2\sqrt{s}+1\right)^{2n} \,e^{\frac{2\sqrt{s}+1}{2\sqrt{s}}\frac{h^2}{n}}\,\times \left[1+ O(n^{-1})\right], \\
\left(\frac{n-\rho_0-h}{n-\rho_0+h}\right)^{h+1} & = & e^{-\frac{2\sqrt{s}+1}{\sqrt{s}}\frac{h(h+1)}{n}}\,\times \left[1+ O(n^{-1})\right]. 
\eea
Then, (\ref{pnn_asym_M}) becomes 
\be
p^{(h)}_{\rM, \, n,n,\,s} \simeq s^{-h}\sqrt{\frac{2}{\pi\sigma^3}}\,\frac{(h+1)^2}{n^{3/2}}\,e^{-\frac{1}{2\sigma } \frac{(h+1)^2}{n}} 
\, \times \left[1+O(n^{-1})\right]
\label{pnn_asym_M2}
\ee
with $\sigma\equiv \frac{\sqrt{s}}{2\sqrt{s}+1}$.

\section{R\'enyi entropy of colorless Fredkin spin chain} 
\label{sec:renyi_F}
\setcounter{equation}{0}
First, let us compute asymptotic behavior of the R\'enyi entropy (\ref{renyi_F}) as $n\to \infty$ for colorless case ($s=1$). 
In the sum $\sum_{h=0}^n\left(p^{(h)}_{\rF,\,n,n,\,s=1}\right)^\alpha$ with (\ref{pnn_asym_F}), 
there is a saddle point $h_*=\sqrt{n}\,(1+O(n^{-1}))$ which solves $\ln \frac{n+h}{n-h}=\frac{2}{h+1}$. 
This justifies use of (\ref{pnn_asym_F2}). 

After converting the sum to an integral: 
\be
\sum_{h=0}^n\frac{1+(-1)^{n+h}}{2}\,(\cdots) \to \frac12\sqrt{n}\int_0^\infty dx\,(\cdots)
\ee
with $x=\frac{h}{\sqrt{n}}$, we obtain 
\be
\sum_{h=0}^n\left(p^{(h)}_{\rF,\,n,n,\,s=1}\right)^\alpha \simeq \left(\frac{8}{\sqrt{\pi}}\right)^\alpha\,\frac12 \,n^{\frac{1-\alpha}{2}}
\left(\int^\infty_0dx\,x^{2\alpha}\,e^{-\alpha x^2}-\int_0^{n^{-1/2}}dx\,x^{2\alpha}\,e^{-\alpha x^2}\right).
\ee
The first integral in the parenthesis is given by $\frac12\,\alpha^{-\alpha-\frac12}\Gamma\left(\alpha+\frac12\right)$.  
The second one is evaluated as $O(n^{-\alpha-\frac12})$, since the exponential in the integrand can be expanded as 
\be
e^{-\alpha x^2} = 1-\alpha x^2 +\cdots = 1+O(n^{-1})
\ee
for $0<x<n^{-1/2}$.  

Hence, the R\'enyi entropy asymptotically behaves as 
\bea
S_{\rF,\,A,\alpha} & = & \frac12 \ln n+\frac{1}{1-\alpha}\ln\Gamma\left(\alpha+\frac12\right) 
-\frac{1}{2(1-\alpha)} \left\{(1+2\alpha)\ln \alpha + \alpha\ln \frac{\pi}{64} +\ln 16\right\}
\nn \\
 & & +(\mbox{terms vanishing as $n\to \infty$}).
 \label{renyi_asym_Ff}
 \eea 
This grows logarithmically as $n \to \infty$, and gives the von Neumann entanglement entropy in the $\alpha\to 1$ limit: 
\be
S_{\rF,\,A}=\frac12\ln n+\frac12\ln \frac{\pi}{4}+\gamma-\frac12 +(\mbox{terms vanishing as $n\to \infty$}).
\label{neumann_asym_F}
\ee
with $\gamma$ being the Euler constant. 
(\ref{neumann_asym_F}) is consistent with the result obtained in \cite{fredkin}.   
 
\section{R\'enyi entropy of colorless Motzkin spin chain}
\label{sec:renyi_M}
\setcounter{equation}{0}
For the R\'enyi entropy (\ref{renyi_M}) with $s=1$, we repeat similar computation to the Fredkin case by using (\ref{pnn_asym_M2}). 
The result is  
\bea
S_{\rM,\,A,\alpha} & = & \frac12 \ln n+\frac{1}{1-\alpha}\ln\Gamma\left(\alpha+\frac12\right) 
-\frac{1}{2(1-\alpha)} \left\{(1+2\alpha)\ln \alpha + \alpha\ln \frac{\pi}{24} +\ln 6\right\}
\nn \\
 & & +(\mbox{terms vanishing as $n\to \infty$}).
 \label{renyi_asym_Mf}
\eea
Again, this grows logarithmically as $n \to \infty$, and reproduces the von Neumann entanglement entropy 
\be
S_{\rM,\,A}=\frac12\ln n+\frac12\ln \frac{2\pi}{3}+\gamma-\frac12 +(\mbox{terms vanishing as $n\to \infty$}).
\label{neumann_asym_M}
\ee
in the $\alpha\to 1$ limit.   
(\ref{renyi_asym_Mf}) reproduces half-chain case in the result obtained in \cite{movassagh}. 

\section{R\'enyi entropy of $s$-color Fredkin spin chain} 
\label{sec:renyi_Fs}
\setcounter{equation}{0}
We compute asymptotic behavior of the R\'enyi entropy (\ref{renyi_F}) for colored case ($s>1$). 
From the expression (\ref{pnn_asym_F}), we find 
\be
\sum_{h=0}^n s^h\left(p^{(h)}_{\rF,\,n,n,\,s}\right)^\alpha \simeq 
\left(\frac{8}{\sqrt{\pi}}\right)^\alpha \,n^{-\frac32\alpha}\,e^{2\alpha(n+2)\ln n} \,
\sum_{h=0}^n\frac{1+(-1)^{n+h}}{2}\,e^{f_{\rF}(h)} \, \times  \left[1+O(n^{-1})\right]  
\label{sum_F}
\ee
with 
\be
f_{\rF}(h) \equiv (1-\alpha)(\ln s) \,h -\alpha (n+h+3)\ln(n+h)-\alpha (n-h+1)\ln(n-h) + 2\alpha\ln (h+1).
\label{fh_F}
\ee

Two cases $0<\alpha <1$ and $\alpha>1$ are separately discussed. 

\subsection{$0<\alpha<1$ Case}
We evaluate the sum (\ref{sum_F}) by the saddle point method for large $n$. The saddle point equation $f_{\rF}(h)=f_{\rF}(h+2)$ becomes 
\be
\frac{1-\alpha}{\alpha}\ln s = \ln\frac{n+h}{n-h} + O(n^{-1}, \,h^{-1})
\ee
for $n, \,h, \,n\pm h\gg 1$. 
This is solved by 
\be
h_*=n\tanh \frac{\theta}{2}+O(n^0) \qquad \mbox{with}\quad \theta\equiv \frac{1-\alpha}{\alpha}\ln s.
\label{h*_F}
\ee
Differently from the colorless case, the saddle point value $h_*$ is of the order $O(n)$. 
Thus, we should use (\ref{pnn_asym_F}) instead of (\ref{pnn_asym_F2}). 

We obtain 
\bea
f_{\rF}(h_*) & = & -(1-\alpha)\ln s -2\alpha (n+2)\ln\frac{n}{\cosh\frac{\theta}{2}} +2\alpha\ln\left(n\tanh\frac{\theta}{2}\right) + O(n^{-1}),  
\nn \\
f_{\rF}''(h_*) & = & -\frac{2\alpha}{n}\cosh^2\frac{\theta}{2}\,\times \left[1+O(n^{-1})\right],   
\label{fh_F2}
\eea
and evaluate the sum as 
\bea
\sum_{h=0}^n\frac{1+(-1)^{n+h}}{2}\,e^{f_{\rF}(h)} 
& = & e^{f_{\rF}(h_*)}\,\frac12\int^\infty_{-\infty}dx \,e^{\frac12f_{\rF}''(h_*)x^2} \times \left[1+O(n^{-1})\right]   \nn \\
& = &  e^{f_{\rF}(h_*)}\,\frac12\sqrt{\frac{2\pi}{-f_{\rF}''(h_*)}}\times \left[1+O(n^{-1})\right].  
\label{sum_F2}
\eea
Since effectively $|x|\leq O(\sqrt{n})$ due to the Gaussian integral, 
higher order term of $x^k$ can be regarded as the order 
\be
\frac{1}{k!}\,f^{(k)}_{\rF}(h_*)\,x^k =O(n^{1-k}) \times O(n^{k/2}) =O(n^{1-k/2}).
\ee
From parity $x\to -x$ in the integral, we can see that higher order terms of $k\geq 3$ merely contribute 
to the factor $\left[1+O(n^{-1})\right]$ in (\ref{sum_F2}). 

Combining (\ref{renyi_F}), (\ref{sum_F}) and (\ref{sum_F2}) with (\ref{fh_F2}) leads to
\bea
S_{\rF,\,A, \alpha} & = & n\frac{2\alpha}{1-\alpha}\,\ln\cosh\frac{\theta}{2} +\frac{1+\alpha}{2(1-\alpha)}\,\ln n \nn \\
& & -\ln s +\frac12\ln\frac{\pi}{4} -\frac{1}{2(1-\alpha)}\,\ln \alpha-\frac{1}{1-\alpha}\,\ln\cosh\frac{\theta}{2}
+\frac{2\alpha}{1-\alpha}\,\ln \sinh\theta \nn \\
& & +(\mbox{terms vanishing as $n\to \infty$})
\label{renyi_asym_F1f}
\eea
with $\theta\equiv \frac{1-\alpha}{\alpha}\ln s$. 
The R\'enyi entropy grows proportionally to the volume as $n\to \infty$. 
Note that we cannot take $\alpha\to 1$ or $s\to 1$ limit in (\ref{renyi_asym_F1f}). 
In the limit, $\theta$ becomes zero and the leading term of $h_*$ vanishes in (\ref{h*_F}), which makes invalid the computation so far. 
Namely, the $n \to \infty$ limit does not commute with $\alpha\to 1$ or $s\to 1$ limit.

\subsection{$\alpha>1$ Case}
In this case, the summand of $\sum_{h=0}^n s^h \left(p^{(h)}_{\rF,\,n,n,\,s}\right)^\alpha$ contains a damping factor $s^{-(\alpha-1)h}$ 
as $h$ grows. 
Due to this, 
$h\lesssim \frac{1}{(\alpha-1)\ln s}=O(n^0)$ dominantly contributes to the sum.  
Hence, we can use (\ref{pnn_asym_F2}): 
\bea
\sum_{h= 0}^n s^h \left(p^{(h)}_{\rF,\,n,n,\,s}\right)^\alpha 
& \simeq & \left(\frac{8}{\sqrt{\pi}}\right)^\alpha\,n^{-\frac32\alpha} \sum_{h=0}^n\frac{1+(-1)^{n+h}}{2}\,s^{-(\alpha-1)h}(h+1)^{2\alpha}\,
e^{-\frac{\alpha}{n}(h+1)^2} \nn \\
& & \times \left[1+O(n^{-1})\right] . 
\eea
Since $h$ can be regarded as $h\leq O(n^0)$ in the sum, we expand the factor $e^{-\frac{\alpha}{n}(h+1)^2}$ as 
\be
e^{-\frac{\alpha}{n}(h+1)^2}=1-\frac{\alpha}{n}(h+1)^2 + \cdots = 1+O(n^{-1}),  
\ee
and find 
\be
\sum_{h=0}^n s^h \left(p^{(h)}_{\rF,\,n,n,\,s}\right)^\alpha 
\simeq \left(\frac{8}{\sqrt{\pi}}\right)^\alpha \,n^{-\frac32\alpha}s^{\alpha-1}\sum_{h\geq 1}\frac{1-(-1)^{n+h}}{2}\,s^{-(\alpha-1)h}h^{2\alpha} 
\,  \times \left[1+O(n^{-1})\right] .
\ee
Since the sum in the r.h.s. does not contain any large or small quantity which is suitable for converting it to an integral, 
we should compute the sum as it is. 
We can change the lower bound of the sum from $h=1$ to 0, because the summand at $h=0$ vanishes. 
In terms of the Lerch transcendent 
\be
\Phi(z,g,a)=\sum_{k=0}^\infty \frac{z^k}{(k+a)^g},
\label{Lerch}
\ee
the sum is recast for $n$ even:
\bea
\sum_{h=0}^n\frac{1-(-1)^{n+h}}{2}\,s^{-(\alpha-1)h}h^{2\alpha} 
& = &  \sum_{k=0}^\infty s^{-(\alpha-1)(2k+1)}\,(2k+1)^{2\alpha} \nn \\
& = & s^{-(\alpha-1)} 2^{2\alpha}\Phi\left(s^{-2(\alpha-1)}, -2\alpha, \frac12\right),
\label{Phi_even}
\eea
and for $n$ odd:
\be
\sum_{h=0}^n\frac{1-(-1)^{n+h}}{2}\,s^{-(\alpha-1)h}h^{2\alpha}  =\sum_{k=0}^\infty s^{-(\alpha-1)2k}\,(2k)^{2\alpha} 
= 2^{2\alpha} \Phi(s^{-2(\alpha-1)}, -2\alpha, 0).
\label{Phi_odd}
\ee

From these results, we find that the R\'enyi entropy takes the form: 
\bea
S_{\rF,\,A, \,\alpha} & =& \frac{3\alpha}{2(\alpha-1)}\,\ln n +\frac{\alpha}{2(\alpha-1)}\,\ln \frac{\pi}{32^2}
-\frac{1}{\alpha-1}\ln \Phi\left(s^{-2(\alpha-1)}, -2\alpha, \frac12\right) \nn \\
& &  +(\mbox{terms vanishing as $n\to \infty$})
\label{renyi_asym_F2e}
\eea
for $n$ even, and 
\bea
S_{\rF,\,A, \,\alpha} & =& \frac{3\alpha}{2(\alpha-1)}\,\ln n -\ln s+\frac{\alpha}{2(\alpha-1)}\,\ln \frac{\pi}{32^2} 
-\frac{1}{\alpha-1}\ln \Phi\left(s^{-2(\alpha-1)}, -2\alpha, 0\right) \nn \\
& & +(\mbox{terms vanishing as $n\to \infty$})
\label{renyi_asym_F2o}
\eea
for $n$ odd.
For both of (\ref{renyi_asym_F2e}) and (\ref{renyi_asym_F2o}), the R\'enyi entropy asymptotically behaves as logarithm of the volume.  
Again we cannot take $\alpha\to 1$ or $s\to 1$ limit in these expressions, because the damping by $s^{-(\alpha-1)h}$ 
ceases in the limit and the computation becomes invalid. 

\begin{figure}[h!]
  \begin{center}
   \includegraphics[width=70mm]{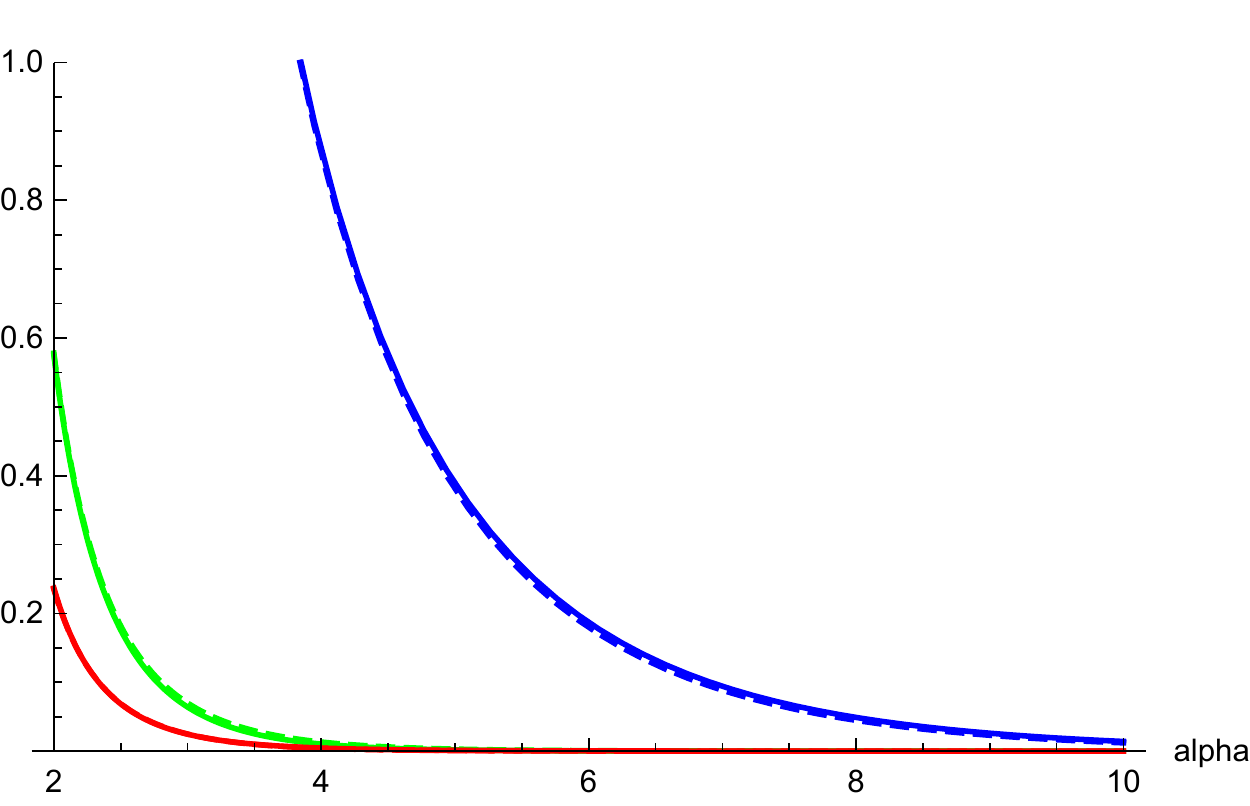}
  \end{center}
  \caption{\small The l.h.s. and r.h.s. of (\ref{LP12}) are plotted from $\alpha=2$ to 10. 
  The l.h.s. (r.h.s.) for $s=2, 4$ and 6 are depicted as blue real (dashed) curve, green real (dashed) curve and red real (dashed) curve. 
  In each $s$, the real and dashed curves are almost on top of each other. }
  \label{fig:LP12}
 \end{figure}

\begin{figure}[h!]
  \begin{center}
   \includegraphics[width=70mm]{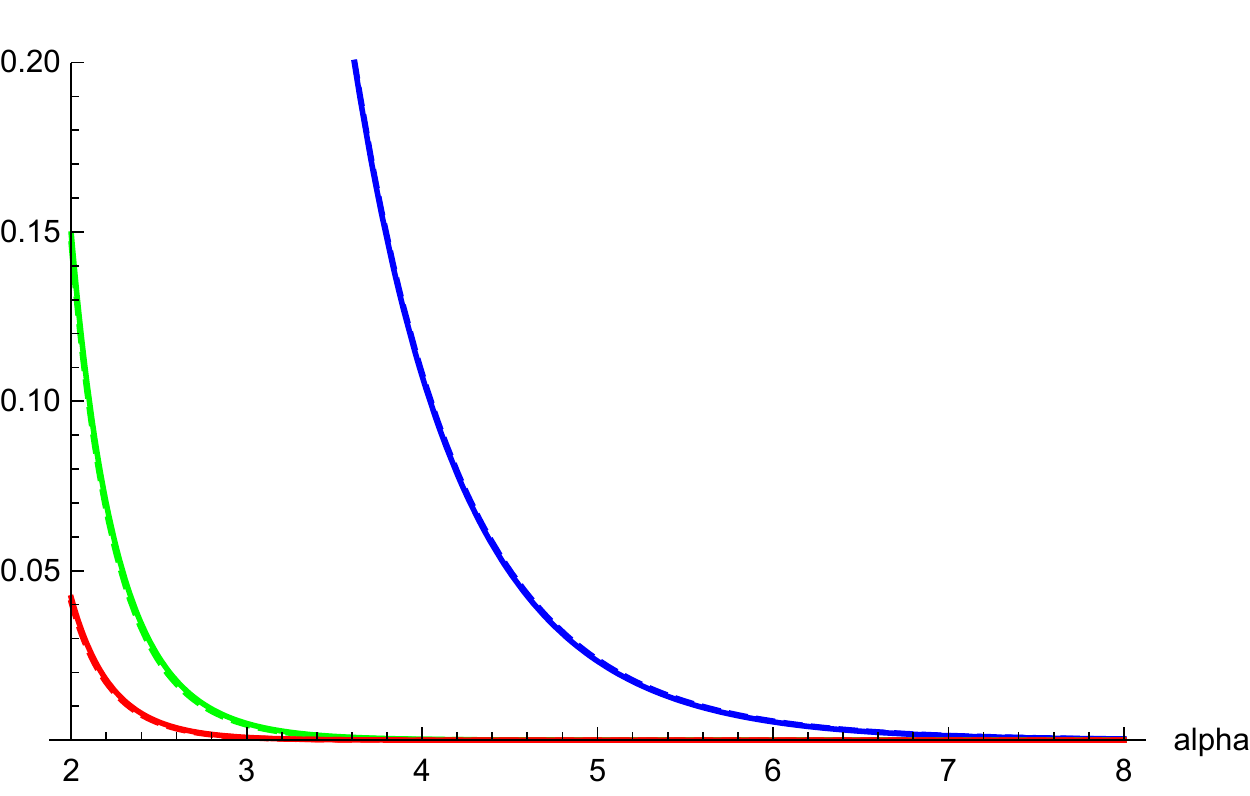}
  \end{center}
  \caption{\small The l.h.s. and r.h.s. of (\ref{LP0}) are plotted from $\alpha=2$ to 8. 
  The l.h.s. (r.h.s.) for $s=2, 4$ and 6 are depicted as blue real (dashed) curve, green real (dashed) curve and red real (dashed) curve. 
  In each $s$, the real and dashed curves are almost on top of each other.}
  \label{fig:LP0}
 \end{figure}

To see qualitative behavior of (\ref{Phi_even}) and (\ref{Phi_odd}), let us evaluate them with the sums replaced by integrals. 
Then, 
\bea
\Phi\left(s^{-2(\alpha-1)}, -2\alpha, \frac12\right)
& \sim & \frac{s^{\alpha-1}\Gamma(2\alpha+1, (\alpha-1)\ln s)}{(2(\alpha-1)\ln s)^{2\alpha+1}} , 
\label{LP12}
\\
\Phi\left(s^{-2(\alpha-1)}, -2\alpha, 0\right)
&\sim & \frac{\Gamma(2\alpha+1)}{(2(\alpha-1)\ln s)^{2\alpha+1}}, 
\label{LP0}
\eea
where $\Gamma(z,x)$ is the incomplete Gamma function:
\be
\Gamma(z,x)\equiv \int^\infty_xdt\,t^{z-1}e^{-t}.
\ee
The l.h.s. and r.h.s. of (\ref{LP12}) and (\ref{LP0}) are plotted in Figs.~\ref{fig:LP12} and \ref{fig:LP0} for $s=2,4,6$. 
In each plot, the difference of the l.h.s. and r.h.s. is almost invisible. 


\subsection{Von Neumann entanglement entropy}
The expression of the von Neumann entanglement entropy (\ref{vonneumann}) becomes
\be
S_{\rF,\,A}=-\sum_{h= 0}^n s^h\,p^{(h)}_{\rF,\,n,n,\,s}\ln p^{(h)}_{\rF,\,n,n,\,s} 
\label{vonneumann_F}
\ee
with (\ref{pnn_asym_F}) or (\ref{pnn_asym_F2}). 
Note that the first factor $s^h$ in the summand cancels with $s^{-h}$ in $p^{(h)}_{\rF,\,n,n,\,s}$, 
and the summand does not contain exponentially growing factor with $h$. 
Then, the saddle point of $h$ in (\ref{vonneumann_F}) is $O(\sqrt{n})$, which allows us to use (\ref{pnn_asym_F2}). 
By converting the sum to an integral with $h=\sqrt{n}\,x$, we have 
\be
S_{\rF,\,A}=-\frac{4}{\sqrt{\pi}}\int_{n^{-1/2}}^{\infty}dx\,x^2e^{-x^2}\,\ln\left[s^{-\sqrt{n}\,x}\frac{8s}{\sqrt{\pi\,n}}\,x^2e^{-x^2}\right]
\,\times\left[1+O(n^{-1})\right].
\ee
We divide the integral $\int_{n^{-1/2}}^\infty$ to $\int_0^\infty - \int_0^{n^{-1/2}}$, where the second integral is evaluated as 
$O\left(\frac{\ln n}{n^{3/2}}\right)$ and can be neglected. 
By computing the first integral, we obtain
\be
S_{\rF,\,A}=(2\ln s)\sqrt{\frac{n}{\pi}} + \frac12\ln n +\frac12\ln \frac{\pi}{4} + \gamma-\frac12-\ln s + 
(\mbox{terms vanishing as $n\to \infty$}).
\label{vonneumann_Ff}
\ee
This is consistent with the result in \cite{fredkin}. 
Note that in the computation we should not approximate $h+1$ to $h$ in (\ref{pnn_asym_F2}) to obtain the $O(n^0)$ terms correctly. 
Since the leading of (\ref{vonneumann_Ff}) is $O(\sqrt{n})$ and $h$ can be regarded as at most $O(\sqrt{n})$ in the Gaussian integral, 
the approximation $h+1\to h$ can affect the $O(n^0)$ terms (actually, $-\ln s$ in the $O(n^0)$ terms disappears).

\subsection{Phase transition}
The R\'enyi entropy $S_{\rF,\,A,\, \alpha}$ exhibits different asymptotic behavior for $0<\alpha<1$ and $\alpha>1$. 
It grows proportional to the volume for the former case, whereas it behaves as logarithm of the volume for the latter case. 
In the definion of the R\'enyi entropy (\ref{renyi}), $H_{{\rm ent},\,A}\equiv -\ln \rho_A$ is called as entanglement Hamiltonian. 
Then, the parameter $\alpha$ is analogous to the inverse temperature.  
From this point of view, our result means that phase transition takes place at the inverse temperature $\alpha=1$.    
As we saw, Dyck walks with large height $h=O(n)$ dominantly contribute to the $S_{\rF,\,A, \,\alpha}$ in 
``high temperature'' region $0<\alpha<1$, 
which leads to the volume law behavior. On the other hand, Dyck walks with low height $h=O(n^0)$ dominate in 
``low temperature'' $\alpha>1$, 
which does not change qualitative behavior of the colorless case. 

The transition point itself forms a phase, where the von Neumann entanglement entropy behaves as a square root of the volume. 
Main contribution to (\ref{vonneumann_Ff}) comes from $h=O(\sqrt{n})$.

\section{R\'enyi entropy of $s$-color Motzkin spin chain}
\label{sec:renyi_Ms}
In this section, we compute large-$n$ behavior of the R\'enyi entropy of $s$-color Motzkin spin chain ($s>1$). 
Plugging (\ref{pnn_asym_M}) to (\ref{renyi_M}), what we should evaluate is
\be
 \sum_{h=0}^n s^h\left(p^{(h)}_{\rM,\,n, n,\,s}\right)^\alpha \simeq \left(\frac{1}{\sqrt{\pi}\,s^{1/4}}\right)^\alpha\,\frac{(2n)^{\frac32\alpha}
\,n^{(2n+1)\alpha}}{\left(2\sqrt{s}+1\right)^{(2n+\frac32)\alpha}} \,\sum_{h= 0}^n e^{f_{\rM}(h)}  \, \times \left[1+ O(n^{-1})\right],
\label{sum_M}
\ee
where 
\bea
f_{\rM}(h) & \equiv & (1-\alpha)(\ln s) h -\alpha (2n+3)\ln \rho_{0} 
 +\alpha(h+1)\ln \frac{n-\rho_{0}-h}{n-\rho_{0}+h} \nn \\
 & & +2\alpha\ln(h+1)-\frac{\alpha}{2}\ln\left[4sn^2-(4s-1)h^2\right],
\label{fh_M} 
\eea
and $\rho_0$ is given by (\ref{saddlepoint_r}).

We discuss the two cases of $0<\alpha<1$ and $\alpha>1$ separately. 

\subsection{$0<\alpha<1$ Case} 
Strategy is the same as in the colored Fredkin case. 
A saddle point with respect to the sum (\ref{sum_M}) is given by the equation $f_{\rM}(h)=f_{\rM}(h+1)$, 
which can be expressed as $f'_{\rM}(h)+O(n^{-1},\,h^{-1})=0$ for $n,\,h,\,n\pm h\gg 1$. 
From the relation $(n-\rho_0)^2-h^2=4s\rho_0^2$, we obtain
\be
\frac{d\rho_{0}}{dh}=-\frac{h}{(4s-1)\rho_{0}+n}=-\frac{\sqrt{(n-\rho_{0})^2-4s\rho_{0}^2}}{(4s-1)\rho_{0}+n}
\ee
and 
\be
f'_{\rM}(h)=(1-\alpha)\ln s -\alpha\ln\frac{1+\sqrt{1-\frac{4s\rho_0^2}{(n-\rho_0)^2}}}{1-\sqrt{1-\frac{4s\rho_0^2}{(n-\rho_0)^2}}}
+O(n^{-1},\,h^{-1}).
\ee
The saddle point equation is solved by
\be
h_* = n\frac{s^{\frac{1}{2\alpha}}-s^{1-\frac{1}{2\alpha}}}{s^{\frac{1}{2\alpha}}+s^{1-\frac{1}{2\alpha}}+1} + O(n^0),
\qquad 
\left. \rho_0\right|_{h=h_*} = \frac{n}{s^{\frac{1}{2\alpha}}+s^{1-\frac{1}{2\alpha}}+1} + O(n^0).
\label{h*_M}
\ee
Again, $h_*=O(n)$ and we should use (\ref{pnn_asym_M}) instead of (\ref{pnn_asym_M2}). 

Then~\footnote{
Note that $O(n^0)$ terms in $h_*$ contribute only to $O(n^{-1})$ terms in (\ref{fh*_M}). 
In writing $h_*$ as $h_* =n\xi+\eta$ with $\xi, \eta=O(n^0)$, 
\be
f_{\rM}(h_*)=f_{\rM}(n\xi) + f'_{\rM}(n\xi)\eta + \frac12f''_{\rM}(n\xi)\eta^2 + \cdots.  
\ee
Since $f'_{\rM}(n\xi)=0 + O(n^{-1})$, $f''_{\rM}(n\xi)=O(n^{-1})$ and so on, we can see $\eta$ affects only $O(n^{-1})$ terms in $f_{\rM}(h_*)$.  
}, 
\bea
f''_{\rM}(h) & = & - \frac{2\alpha n}{\rho_0}\,\frac{1}{n-\rho_0+4s\rho_0} \,\times \left[1+ O(n^{-1})\right], \\
f''_{\rM}(h_*) & = & -\frac{2\alpha}{n}\,\frac{\left(s^{\frac{1}{2\alpha}}+s^{1-\frac{1}{2\alpha}}+1\right)^2}{s^{\frac{1}{2\alpha}}+s^{1-\frac{1}{2\alpha}}+4s}
\,\times \left[1+ O(n^{-1})\right],\\
f_{\rM}(h_*) & = & -(2n+2)\alpha\ln n+(2n+3)\alpha \ln \left(s^{\frac{1}{2\alpha}}+s^{1-\frac{1}{2\alpha}}+1\right) -(1-\alpha)\ln s \nn \\
& & -\frac{\alpha}{2}\ln\left[1+4\frac{\left(2s^{\frac{1}{2\alpha}}+1\right)\,\left(2s^{1-\frac{1}{2\alpha}}+1\right)}{\left(s^{\frac{1-\alpha}{2\alpha}}- s^{-\frac{1-\alpha}{2\alpha}} \right)^2}\right] +O(n^{-1}).
\label{fh*_M}
\eea

Evaluating the sum (\ref{sum_M}) in the saddle point method: 
\bea
\sum_{h= 0}^n e^{f_{\rM}(h)} 
& = & e^{f_{\rM}(h_*)}\,\frac12\int^\infty_{-\infty}dx \,e^{\frac12f_{\rM}''(h_*)x^2} \times \left[1+O(n^{-1})\right]   \nn \\
& = &  e^{f_{\rM}(h_*)}\,\frac12\sqrt{\frac{2\pi}{-f_{\rM}''(h_*)}}\times \left[1+O(n^{-1})\right],
\label{sum_M2}
\eea
we eventually find 
\bea
S_{\rM,\,A, \,\alpha} & = & \frac{1}{1-\alpha} \ln \sum_{h=0}^n s^h\left(p^{(h)}_{n,n,\,s}\right)^\alpha \nn \\ 
& =& n \frac{2\alpha}{1-\alpha}\,\ln\left[\sigma\left(s^{\frac{1-\alpha}{2\alpha}}+s^{-\frac{1-\alpha}{2\alpha}}+s^{-1/2}\right)\right]+\frac{1+\alpha}{2(1-\alpha)}\,\ln n \nn \\
& & +\frac12\ln \pi -\frac{1}{1-\alpha}\ln\left(s\sqrt{\alpha}\right) +\frac{1}{2(1-\alpha)}\,\ln\left(s^{\frac{1}{2\alpha}}+s^{1-\frac{1}{2\alpha}}+4s\right)  +\frac{3\alpha}{2(1-\alpha)}\,\ln(2\sigma)  \nn \\
& & +\frac{3\alpha-1}{1-\alpha}\,\ln\left(s^{\frac{1}{2\alpha}}+s^{1-\frac{1}{2\alpha}}+1\right)
 -\frac{\alpha}{2(1-\alpha)}\,\ln \left[1+4\frac{\left(2s^{\frac{1}{2\alpha}}+1\right)\,\left(2s^{1-\frac{1}{2\alpha}}+1\right)}{\left(s^{\frac{1-\alpha}{2\alpha}}- s^{-\frac{1-\alpha}{2\alpha}} \right)^2}\right] \nn \\
& & +(\mbox{terms vanishing as $n\to \infty$}) 
\label{renyi_asym_M1f}
\eea
with $\sigma =\frac{\sqrt{s}}{2\sqrt{s}+1}$. 
The R\'enyi entropy has asymptotic behavior proportional to the volume, which is similar to what we saw in the $s$-color Fredkin case 
(\ref{renyi_asym_F1f}). 
The coefficient of $\ln n$ term coincides with that in the colored Fredkin case (\ref{renyi_asym_F1f}), which seems to show some universal 
property.

\subsection{$\alpha>1$ Case} 
As in the colored Fredkin case, 
due to the exponential damping factor $s^{-(\alpha-1)h}$ in the sum $\sum_{h=0}^n s^h\left(p^{(h)}_{\rM,\,n,n,\,s}\right)^\alpha$, 
we can regard $h$ as a quantity at most $O(n^0)$, which justifies use of (\ref{pnn_asym_M2}). 

The sum is recast as 
\bea
 & & \sum_{h=0}^n s^h\left(p^{(h)}_{\rM,\,n,n,\,s}\right)^\alpha \simeq 
\left(\frac{4}{\sqrt{\pi}}\right)^\alpha\,(2\sigma n)^{-\frac32\alpha} \,s^{\alpha-1} 
\sum_{h\geq 0}s^{-(\alpha-1)h}\,h^{2\alpha}\,e^{-\frac{\alpha}{2\sigma n}h^2} \nn \\
& & \hspace{7mm}= \left(\frac{4}{\sqrt{\pi}}\right)^\alpha\,(2\sigma n)^{-\frac32\alpha} \,s^{\alpha-1} \Phi\left(s^{-(\alpha-1)}, -2\alpha, 0\right) 
\,\times \left[1+ O(n^{-1})\right] .
\eea
where the last factor in the first line $e^{-\frac{\alpha}{2\sigma n}h^2}$ can be regarded as $1+O(n^{-1})$, 
and the sum is expressed by the Lerch transcendent (\ref{Lerch}). 
Thus, asymptotic behavior of the R\'enyi entropy is found to increase as $\ln n$: 
\bea
S_{\rM,\,A,\,\alpha} & = & \frac{3\alpha}{2(\alpha-1)}\,\ln n -\ln s+\frac{3\alpha}{2(\alpha-1)}\,\ln\sigma 
+ \frac{\alpha}{2(\alpha-1)}\,\ln \frac{\pi}{2} \nn \\
& & \hspace{-7mm}  -\frac{1}{\alpha-1}\,\ln \Phi\left(s^{-(\alpha-1)}, -2\alpha, 0\right) 
+(\mbox{terms vanishing as $n\to \infty$}) .
\label{renyi_asym_M2f}
\eea
The coefficient of $\ln n$ coincides with that in the colored Fredkin case (\ref{renyi_asym_F2e}) and (\ref{renyi_asym_F2o}), 
which seems to show some universal meaning. 
Qualitative behavior of $\Phi\left(s^{-(\alpha-1)}, -2\alpha, 0\right)$ is evaluated as (\ref{LP0}) with $s$ replaced by $s^{1/2}$. 

For both of the expressions (\ref{renyi_asym_M1f}) and (\ref{renyi_asym_M2f}), we cannot take $\alpha\to 1$ or $s\to 1$ limit, 
since the $n\to \infty$ limit does not commute with $\alpha\to 1$ or $s\to 1$. 
Note that the $O(n)$ term of $h_*$ in (\ref{h*_M}) vanishes in the limit.

\subsection{Von Neumann entanglement entropy}
Similar to the colored Fredkin case, the von Neumann entanglement entropy (\ref{vonneumann}) becomes
\be
S_{\rM,\,A}=-\sum_{h=0}^n s^h\,p^{(h)}_{\rM,\,n,n,\,s}\ln p^{(h)}_{\rM,\,n,n,\,s} 
\label{vonneumann_M}
\ee
with (\ref{pnn_asym_M}) or (\ref{pnn_asym_M2}). 
We can use (\ref{pnn_asym_M2}), because the saddle point of $h$ of the sum (\ref{vonneumann_M}) is $O(\sqrt{n})$.  
After similar computation to the Fredkin case, we obtain
\be
S_{\rM,\,A}=(2\ln s)\sqrt{\frac{2\sigma n}{\pi}} + \frac12\ln n +\frac12\ln (2\pi\sigma) + \gamma-\frac12-\ln s + 
(\mbox{terms vanishing as $n\to \infty$}).
\label{vonneumann_Mf}
\ee
This reproduces the result in \cite{motzkin} except the last term $-\ln s$ of the order $O(n^0)$. 
Again, in the computation we should not approximate $h+1$ to $h$ in (\ref{pnn_asym_M2}) to obtain the $O(n^0)$ terms correctly. 
The approximation $h+1\to h$ amounts to lose the last term $-\ln s$~\footnote{
We think that this is the reason of discrepancy between the analytic result and the numerical 
result in Fig. 2 in SI of \cite{motzkin}.
}. 

\subsection{Phase transition}
Similar to the $s$-colored Fredkin case, 
the R\'enyi entropy $S_{\rM,\,A,\, \alpha}$ has different asymptotic behavior for $0<\alpha<1$ and $\alpha>1$ -- 
linear of the volume and logarithm of the volume.
Motzkin walks with large height $h=O(n)$ dominantly contribute to the $S_{\rM,\,A, \,\alpha}$ in ``high temperature'' region $0<\alpha<1$, 
leading to the volume law behavior, whereas Motzkin walks with low height $h=O(n^0)$ dominate in ``low temperature'' $\alpha>1$, 
qualitative same as the colorless case. 

The transition point itself consists of a phase, where the von Neumann entanglement entropy behaves as a square root of the volume. 
Main contribution to (\ref{vonneumann_Mf}) comes from height $h=O(\sqrt{n})$.



\begin{thebibliography}{99}
\addtolength{\parskip}{-1ex}

\bibitem{bennett_etal}
C.H.~Bennett, H.J.~Bernstein, S.~Popescu, B.~Schumacher, ``Concentrating partial
 entanglement by local operations'',
{\it Phys.\ Rev.}  {\bf A53}, 2046--2052 (1996).

\bibitem{nielsen-chuang}
M.A.~Nielsen, I.L.~Chuang, {\it Quantum Information and Quantum Computation} 
\newblock (Cambridge Univ Press, Cambridge, UK, 2010), 10th anniversary edition.

\bibitem{witten}
E.~Witten, ``Notes on Some Entanglement Properties of Quantum Field Theory'',
arXiv:1803.04993 [hep-th].
  
\bibitem{renyi_book}
A.~R\'enyi, {\it Probability Theory}
\newblock (North Holland, Amsterdam, 1970).

\bibitem{eisert_Rev}
J.~Eisert, M.~Cramer, M.B.~Plenio, ``Colloquium: Area laws for the entanglement entropy'',
{\it Rev.\ Mod.\ Phys.} {\bf 82}, 277  (2010) and 
arXiv:arXiv:0808.3773 [quant-ph].

\bibitem{AKLT1}
L.~Affleck, T.~Kennedy, E.H.~Lieb, H.~Tasaki, ``Valence bond ground states in isotropic quantum antiferromagnets'',
{\it Commun.\ Math.\ Phys.} {\bf 115}, 477--528 (1988).

\bibitem{AKLT2}
I.~Affleck, T.~Kennedy, E.H.~Lieb, H.~Tasaki, ``Rigorous results on valence-bond ground states in antiferromagnets'',
{\it Phys.\ Rev.\ Lett.} {\bf 59}, 799--802 (1987).

\bibitem{fan}
H.~Fan, V.~Korepin, V.~Roychowdhury, ``Entanglement in a valence-bond solid state'',
{\it Phys.\ Rev.\ Lett.} {\bf 93}, 227203 (2004) 
and arXiv:quant-ph/0406067.

\bibitem{vidal}
G.~Vidal, J.~Latorre, E.~Rico, A.~Kitaev, ``Entanglement in quantum critical phenomena'',
{\it Phys.\ Rev.\ Lett.} {\bf 90}, 227902 (2003) and 
arXiv:quant-ph/0211074.

\bibitem{jin}
B.Q.~Jin, V.E.~Korepin, ``Quantum spin chain, Toeplitz determinants and Fisher-Hartwig conjecture'',
{\it J.\ Stat.\ Phys.} {\bf 116}, 79 (2004) and 
arXiv:quant-ph/0304108.

\bibitem{wilczek}
C.~Holzhey, F.~Larsen, F.~Wilczek, ``Geometric and renormalized entropy in conformal field theory'',
{\it Nucl.\ Phys.} {\bf B [FS] 424}, 443--467 (1994).

\bibitem{korepin}
V.E.~Korepin, ``Universality of Entropy Scaling in One Dimensional Gapless Models'', 
{\it Phys.\  Rev.\ Lett.} {\bf 92}, 096402 (2004) and 
arXiv:cond-mat/0311056.
  
\bibitem{calabrese}
P.~Calabrese, J.~Cardy, ``Entanglement entropy and conformal field theory'',
{\it J.\ Phys.} {\bf A42}, 504005 (2009) and 
arXiv:0905.4013 [cond-mat,stat-mech].

\bibitem{wolf}
M.M.~Wolf, ``Violation of the entropic area law for fermions'',
{\it Phys.\ Rev.\ Lett.} {\bf 96}, 010404 (2006).

\bibitem{hastings}
M.B.~Hastings, ``An area law for one-dimensional systems'',
{\it J.\ Stat. Mech. Theory Exp.} P08024 (2007) and 
arXiv:0705.2024 [quant-ph].

\bibitem{huang}
Y.~Huang, (2015) ``Classical simulation of quantum many-body systems'', 
{\it UC Berkeley Electronic Theses and Dissertations} 
https://escholarship.org/uc/item/3ct7d8tf

\bibitem{motzkin}
R.~Movassagh, P.W.~Shor, ``Supercritical entanglement in local systems:
  Counterexample to the area law for quantum matter'',
{\it Proc.\ Natl.\ Acad.\ Sci.} {\bf 113}, 13278--13282 (2016) and 
arXiv:1408.1657 [quant-ph].

\bibitem{bravyi_etal}
S.~Bravyi, L.~Caha, R.~Movassagh, D.~Nagaj, P.W.~Shor, ``Criticality without frustration for quantum spin-1 chains'',
{\it Phys.\ Rev.\ Lett.} {\bf 109}, 207202 (2012) and 
arXiv:1203.5801 [quant-ph].

\bibitem{fredkin}
O.~Salberger, V.~Korepin, ``Entangled spin chain'',
{\it Rev.\ Math.\ Phys.} {\bf 29}, 1750031 (2017) 
and arXiv:1605.03842 [quant-ph].

\bibitem{dellanna_etal}
L.~Dell'Anna, O.~Salberger, L.~Barbiero, A.~Trombettoni, V.E.~Korepin, 
``Violation of cluster decomposition and absence of light-cones in local integer and half-integer spin chains'', 
{\it Phys.\ Rev.} {\bf B 94}, 155140 (2016) and 
arXiv:1604.08281 [cond-mat.str-el]. 

\bibitem{alcarez}
F.C.~Alcaraz, P.~Pyatov, V.~Rittenberg,
``Density profiles in the raise and peel model with and without a wall. Physics and combinatorics'',
{\it J.\ Stat. Mech. Theory Exp.} P01006 (2008) and 
arXiv:0709.4575 [cond-mat.stat-mech].
	
\bibitem{chico}
F.C.~Alcaraz, V.~Rittenberg,
``Nonlocal growth processes and conformal invariance'',
{\it J.\ Stat. Mech. Theory Exp.} P05022 (2012) and 
arXiv:1204.1001 [cond-mat.stat-mech].

\bibitem{zhang_k}
Z.~Zhang, I.~Klich, ``Entropy, gap and a multi-parameter deformation of the Fradkin spin chain'',
{\it J. Phys.}  {\bf A50}, 425201 (2017) and 
arXiv:1702.03581 [cond-mat,stat-mech].

\bibitem{salberger_etal}
O.~Salberger, T.~Udagawa, Z.~Zhang, H.~Katsura, I.~Klich, V.~Korepin, 
``Deformed Fredkin spin chain with extensive entanglement'',
{\it J.\ Stat.\ Mech.\ Theory Exp.} {\bf 1706}, 063103 (2017) and 
arXiv:1611.04983 [cond-mat,stat-mech].

\bibitem{zhang_ak}
Z.~Zhang, A.~Ahmadain, I.~Klich, ``Novel quantum phase transition from bounded to extensive entanglement'',
{\it Proc.\ Natl.\ Acad.\ Sci.} {\bf 114}, 5142--5146 (2017) and 
arXiv:1606.07795 [quant-ph].

\bibitem{udagawa_k}
T.~Udagawa, H.~Katsura, ``Finite-size gap, magnetization, and entanglement of deformed Fredkin spin chain'',
{\it J.\ Phys.} {\bf A50}, 405002 (2017) and 
arXiv:1701.00346 [cond-mat.stat-mech].

\bibitem{barbiero_etal}
L.~Brabiero, L.~Dell'Anna, A.~Trombettoni, V.E.~Korepin, ``Haldane topological orders in Motzkin spin chains'',
{\it Phys.\ Rev.} {\bf B96}, 180404 (2017) and 
arXiv:1701.05878 [cond-mat.stat-mech].

\bibitem{SISmotzkin}
F.~Sugino, P.~Padmanabhan, ``Area law violations and quantum phase transitions in modified Motzkin walk spin chains'',
{\it J.\ Stat.\ Mech.\ Theory Exp.} {\bf 1801}, 013101 (2018) and 
arXiv:1710.10426 [quant-ph].

\bibitem{SISfredkin}
P.~Padmanabhan, F.~Sugino, V.~Korepin, ``Quantum phase transitions and localization in semigroup Fredkin spin chain'', 
 arXiv:1804.00978 [quant-ph].
 
\bibitem{caha_n}
L.~Caha, D.~Nagaj, ``The Pair-Flip model: a very entangled translationally invariant spin-chain'', 
 arXiv:1805.07168 [quant-ph].

\bibitem{movassagh}
R.~Movassagh, ``Entanglement and correlation functions of the quantum Motzkin spin-chain'',
{\it J.\ Math.\ Phys.} {\bf 58}, 031901 (2017) and 
arXiv:1602.07761 [quant-ph].

\bibitem{stephan}
J-M.~St\'ephan, G.~Misguich, V.~Pasquier, ``Phase transition in the R\'enyi-Shannon entropy of Luttinger liquids", 
{\it Phys.\ Rev.} {\bf B84}, 195128 (2011) and 
arXiv:1104.2544  [cond-mat.str-el]. 

\bibitem{belin} 
A.~Belin, A.~Maloney, S.~Matsuura, ``Holographic Phases of Renyi Entropies'', 
{\it J.\ High Energy Phys.} {\bf 1312}, 050 (2013) and 
arXiv:1306.2640 [hep-th].


\bibitem{ying_etal}
X.~Ying, H.~Katsura, T.~Hirano, V.E.~Korepin, ``Entanglement and density matrix of a block of spins in AKLT model'',
{\it J.\ Stat.\ Phys.} {\bf 133}, 347--377 (2008) and 
arXiv:0802.3221 [quant-ph].

\bibitem{lieb_sm}
E.~Lieb, T.~Schultz, D.~Mattis, ``Two soluble models of an antiferromagnetic chain'',
{\it Ann.\ Phys.} {\bf 16}, 407--466 (1961).

\bibitem{barouch_md}
E.~Barouch, B.M.~McCoy, M.~Dresden, (1970) ``Statistical mechanics of the XY model. I'',
{\it Phys.\ Rev.} {\bf A2}, 1075--1092 (1970).

\bibitem{barouch_m}
E.~Barouch, B.M.~McCoy, ``Statistical mechanics of the XY model. II. spin-correlation functions'', 
{\it Phys.\ Rev.} {\bf A3}, 786--804 (1971).

\bibitem{franchini_ik}
F.~Franchini, A.~Its, V.~Korepin, ``Renyi entropy of the XY spin chain'', 
{\it J.\ Phys.} {\bf A41}, 025302 (2008) and 
arXiv:0707.2534 [quant-ph].

\bibitem{its_jk}
F.~Franchini, A.~Its, V.~Korepin, ``Entanglement in the XY spin chain'',
{\it J.\ Phys.} {\bf A38}, 2975--2990 (2005) and 
arXiv:quant-ph/0409027.

\bibitem{peschel}
L.~Peschel, ``On the entanglement entropy for an XY spin chain'',
{\it J.\ Stat.\ Mech.\ Theory Exp.} P12005 (2004) and 
arXiv:cond-mat/0410416.

\bibitem{perlmutter}
E.~Perlmutter, ``A universal feature of CFT R\'enyi entropy'',
{\it J.\ High Energy Phys.} {\bf 03}, 117 (2014) and 
arXiv:1308.1083 [hep-th].

\bibitem{chen}
X.~Chen, E.~Fradkin, W.~Witczak-Krempa, ``Gapless quantum spin chains: multiple dynamics and conformal wavefunctions", 
{\it J.\ Phys.} {\bf A50}, 464002 (2017) and 
arXiv:1707.02317 [cond-mat.str-el]. 

\end{thebibliography}
\end{document}